\documentstyle[12pt,aaspp4]{article}

\def\yrm1m{{\rm yr}^{-1}}
\def\yrm1{yr${}^{-1}$}

\begin{document}

\title{BVI time series data of the Galactic Globular Cluster NGC~3201. 
I. RR Lyrae stars\footnote{Based on observations collected at the 
European Southern Observatory, La Silla, Chile.}}

\author{A. M. Piersimoni\altaffilmark{1}, G. Bono\altaffilmark{2}, 
V. Ripepi\altaffilmark{3}}

\affil{1. Osservatorio Astronomico di Collurania, Via M. Maggini,
64100 Teramo, Italy; piersimoni@te.astro.it}
\affil{2. Osservatorio Astronomico di Roma, Via Frascati 33,
00040 Monte Porzio Catone, Italy; bono@mporzio.astro.it}
\affil{3. Osservatorio Astronomico di Capodimonte, Via Moiariello 16,
80131 Napoli, Italy; ripepi@na.astro.it}

\date{02/18/02}

\pagebreak
\begin{abstract} 

We present Johnson BV, and Kron-Cousins I-band time series data 
collected over three consecutive nights in a region of 13
arcmin$^2$ centered on the Galactic Globular Cluster (GGC) NGC~3201. 
The time sampling of current CCD data allowed us to derive accurate 
light curves, and in turn mean magnitudes and colors for a sample of 
53 RR Lyrae. To overcome the thorny problem of differential reddening 
affecting this cluster, we derived new empirical relations connecting 
the intrinsic (B-V) and (V-I) colors of fundamental ($RR_{ab}$) RR Lyrae 
to the luminosity amplitude, the metallicity, and the pulsation period.   
The key features of these relations are the following: {\em i)} they 
rely on stellar parameters which are not affected by reddening; 
{\em ii)} they supply accurate estimates of intrinsic colors across 
the fundamental instability strip and cover a wide metallicity 
range; {\em iii)} they were derived by neglecting the RR Lyrae that 
are affected by amplitude modulation. Moreover, the 
zero-point of the E(B-V) reddening scale was empirically checked  
using the large sample of RR Lyrae in M~3 (Corwin \& Carney 2001), 
a GGC affected by a vanishing reddening.  

According to these relations we estimated individual reddenings for 
RR Lyrae in our sample and the main results we found are the following: 
{\em i)} the mean cluster reddening based on E(B-V) color excesses  
is $<E(B-V)>=0.30\pm0.03$. This estimate is slightly higher than the 
mean reddening evaluations available in the literature or based  
on the dust infrared map by Schlegel, Finkbeiner, \& Davis (1998), i.e. 
$<E(B-V)>=0.26\pm0.02$. Note that the angular resolution of this map 
is $\approx 6$ arcmin, whereas for current reddening map it is $\approx 1$ 
arcmin. {\em ii)} The mean cluster reddening based on E(V-I) color 
excesses is $<E(V-I)>=0.36\pm0.05$. This estimate is only
marginally in agreement with the mean cluster reddening obtained 
using the reddening map by von Braun \& Mateo (2001) and derived 
by adopting cluster Turn-Off stars, i.e. $<E(V-I)>=0.25\pm0.07$.  
On the other hand, current intrinsic spread among individual 
reddenings ($\approx 0.2$ mag) agrees quite well with the 
estimate provided by previous authors. It is noteworthy that 
previous mean cluster reddenings are in very good agreement with  
values obtained using the empirical relations for intrinsic 
RR Lyrae colors provided by Kovacs \& Walker (2001). {\em iii)} 
According to current individual E(B-V) and E(V-I) reddenings and 
theoretical predictions for Horizontal-Branch stars, we found that 
the true distance modulus for this cluster is $13.32\pm0.06$ mag.  
This determination is somehow supported by the comparison between 
predicted and empirical pulsation amplitudes. 
{\em iv)} The comparison between present luminosity amplitudes and 
estimates available in the literature discloses that approximately  
30\% of fundamental RR Lyrae are affected by amplitude modulation 
({\em Blazhko effect}). This finding confirms an empirical evidence 
originally brought out by Szeidl (1976) and by Smith (1981).  
\end{abstract}


\pagebreak

\section{Introduction}

The stellar content of GGCs plays a crucial role in understanding the 
evolutionary properties of old, low-mass stars. 
Even though pioneering observational 
investigations on these fascinating objects appeared more than 50 years ago, 
complete and homogeneous Color-Magnitude Diagrams (CMDs) for both bright 
and very faint stars became available only recently in the literature.  
The accuracy of photometric data is a key ingredient for providing 
reliable estimates of observables to be compared with evolutionary 
predictions. A paramount observational effort has been 
devoted to Horizontal Branch (HB) stars and in particular to RR Lyrae 
variables, since these objects are fundamental primary distance 
indicators in the Galaxy and in Local Group galaxies. 
Moreover and even more importantly, the RR Lyrae distance scale is 
often used for estimating the Turn-Off (TO) luminosity, and in turn 
the age of GGCs (Vandenberg, Stetson, \& Bolte 1996; Cassisi et al. 1998; 
Caputo 1999). This means that cluster data relying on the same 
photometric zero-point can supply more robust age determinations
(Rosenberg et al. 1999,2000). 

During the last few years, several thorough investigations were aimed 
at improving the accuracy of both evolutionary and pulsational 
predictions. New sets of full amplitude, nonlinear RR Lyrae models 
have been constructed which include a nonlocal, time-dependent 
treatment of convective transport, and in turn a self-consistent 
approach to the coupling between pulsation and convection 
(Bono \& Stellingwerf 1994; Bono et al. 1997a; Feuchtinger 1999, 
and references therein). 
In contrast with linear and nonlinear radiative models this new 
approach provided homogeneous sets of theoretical observables 
-modal stability, pulsational amplitudes- for both fundamental 
and first overtone pulsators to be compared with actual 
properties of RR Lyrae stars. Moreover, the pulsation models, 
to account for the behavior of RR Lyrae stars in GGCs and in the 
Galactic field, were constructed by adopting stellar masses and 
luminosities predicted by evolutionary models over a wide metallicity 
range ($0.0001 \le Z \le 0.04$, Bono et al. 1997a,b,c). 
At the same time, evolutionary models experienced substantial improvements 
in the input physics, such as radiative opacities and equation of state
(D'Antona, Caloi, \& Mazzitelli 1997; Vandenberg \& Irwin 1997; 
Cassisi et al. 1998) and the inclusion of atomic diffusion 
(Cassisi et al. 1999). 
The reader interested in a detailed discussion on the impact of these 
new physical ingredients on theoretical observables is referred to the 
comprehensive reviews  by Vandenberg et al. (1996), Caputo (1999), and 
by Castellani (1999).  

On the other hand, several recent observational investigations performed 
by adopting CCD cameras on relatively small telescopes disclosed that 
current samples of variable stars in GGCs are far from being complete.  
In fact, a large accumulation of evidence suggests that CCD observations  
of the innermost cluster regions allowed the identification of 
sizable samples of binary systems (Albrow et al. 2001; Kaluzny 
\& Thompson 2001), exotic objects (Edmonds et al. 2002), oscillating 
blue stragglers (Kaluzny, Olech, \& Stanek 2001, and references therein), 
as well as a substantial increase in the sample of cluster RR Lyrae 
variables (Walker \& Nemec 1996; Walker 1998; Caputo et al. 1999; 
Olech et al. 2001). 
Finally, we mention that new data reduction techniques such as the
image subtraction method (Alard 1999,2000) supply when compared with
profile fitting methods not only a substantial increase in the number
of variable stars detected in crowded regions (Olech et al. 1999) but
also an improvement in the photometric accuracy of light curves. 

Obviously, the accuracy of astrophysical parameters, such as stellar 
masses and luminosities based on the comparison between theory and 
observations, does depend not only on the accuracy of empirical data,  
but also on the size of individual samples. 
As a consequence, GGCs that present a relatively large number of 
RR Lyrae variables can provide tight constraints on 
the evolutionary and pulsational behavior of these objects. 
Keeping in mind this {\em caveat}, we collected new multiband 
time series CCD data of the GGC NGC~3201.   
The main reason why we selected this cluster is that it will 
supply a comprehensive analysis of a large sample of RR Lyrae  
($N_{RR} = 77$, Clement et al. 2001) in an Oosterhoff 
type I (Oo I) cluster. Moreover, up to now only photographic 
light curves of RR Lyrae in NGC~3201 have been available in the 
literature (Cacciari 1984, hereinafter C84).

This cluster is located close to the Galactic plane ($l_{II}=$277.228, 
$b_{II}=$8.641) and it is a valuable target for spectroscopic surveys 
due to its proximity (DM$\approx13.32$ mag), and to its relatively 
low central concentration ($\log\rho_0=2.69 \,\,[L_\odot/pc^3]$, 
Trager et al. 1993). 
A further interesting feature of NGC 3201 is that it presents a 
retrograde orbit (van den Bergh 1993), thus suggesting that it is 
not a typical member of the halo GC population.
The main drawbacks of NGC~3201 are that it is affected by field 
contamination, presents a relatively high reddening 
($<E(B-V)>\approx 0.25-0.30$), and is also affected by differential reddening. 
The most recent CCD photometric studies of this cluster have been provided 
by Alcaino, Liller, \& Alvarado (1989, BVRI bands), Brewer et al. (1993, 
UBV bands), Covino \& Ortolani (1997, BV data, hereinafter CO97).
More recently Rosenberg et al. (2000) and von Braun \& Mateo (2001, 
hereinafter vBM01) collected deep and accurate VI data.   

This is the first paper of a series devoted to the evolutionary and
pulsation properties of stellar population in NGC~3201. 
In \S 2 we present the observations and the strategy adopted to reduce 
the data together with the calibration of the photometric zero-point 
and the comparison with previous investigations available in the literature. 
A brief discussion of the most recent metallicity determinations of 
NGC~3201 is presented in \S 3. 
The pulsational parameters and the light curves of fundamental  
and first overtone (FO) RR Lyrae stars are discussed in \S 4, while 
the subsection 4.1 deals with variables showing amplitude 
modulation ({\em Blazhko effect}). 
In \S 5 we present a detailed analysis of individual reddening 
evaluations for RR Lyrae stars according to (B-V) and (V-I) colors 
(\S 5.1, and 5.2), as well as to the Fourier parameters of V band 
light curves (\S 5.3). In particular, in the first two 
subsections we discuss three new empirical relations connecting 
the intrinsic color of $RR_{ab}$ stars with luminosity amplitude, 
pulsation period, and metallicity.  The comparison between these 
relations and similar relations available in the literature are 
also described in this section. 
The comparison of predicted pulsation properties, instability 
strip and pulsation amplitudes, with empirical data is detailed 
in \S 6. In this section, we briefly discuss the comparison of 
RR Lyrae in NGC~3201 with the sample of RR Lyrae in IC~4499 
and in M~5. The main feature of the CMD are outlined in \S 7. 
In this section, we also present the (V,B-V) and the (V,V-I) 
CMDs dereddened by means of the reddening map derived using 
current sample of $RR_{ab}$ stars.         
The main results of this investigation are summarized in \S 8, 
together with the future developments of this project. Finally, 
in Appendix A, we list a series of comments on individual variables.

\section{Observations and Data Reduction.}

Multiband photometric data were collected with the 1.5m Danish telescope at 
ESO-La Silla equipped with a Loral CCD ($2024\times2024$ pix) during an 
observing run from 21st through to 23rd February 1996. The field of view 
of the CCD is $13\times13$ arcmin and NGC~3201 was approximately 
centered on this field. The exposure times were 180s for the B 
band, and 40s for both V and I bands respectively. We roughly collected 
80 frames per band, and the stars were identified and measured by using 
Daophot and Allstar packages (Stetson 1987, 1995). The internal accuracy 
of our photometry is of the order of hundredths of magnitude at the 
typical magnitude of RR Lyrae stars. However, the sensitivity of the 
CCD drastically decreases in the 300 pixels ($\approx$ 1.9 arcmin) 
close to the edges. As a consequence, the S/N ratio of the stars located 
in this region is significantly lower and the light curves of RR Lyrae 
variables such as V7, V48, V51, and V71 present a larger scatter.  
   
Oddly enough, the absolute calibration of current data has been a 
pleasant experience, since Stetson 
(2000\footnote{Data available at the following Web site {\tt http://cadcwww.hia.nrc.ca/standards}.}) collected a 
sizable sample of local standards across NGC~3201. On the basis 
of these new standards it has been 
possible to derive an accurate calibration, since V and I standards 
(145 stars) cover a region of $4.5 \times 4.5$ arcmin around the center 
of the cluster, while the standards in B (33 stars) are distributed over 
a region of $\sim 2.6 \times 2.6$ arcmin.
Moreover, it is noteworthy that the (B-V) color of the standards range 
from 0.35 to 1.55, while the (V-I) colors range from 0.2 to 1.7. 
This means that current calibration properly covers both blue/extreme 
HB stars as well as stars close to the tip of the Red Giant Branch (RGB).  
Fig. 1 shows the calibration equations we derived according to the 
difference between current instrumental magnitudes and the Stetson's 
standards. On the basis of this comparison we estimate that our 
calibration errors are of the order of 0.02 mag in V and B bands 
and of 0.03 mag in the I band. The uncertainty in the I band is larger 
than in the B,V bands. The difference might be due to the limited 
accuracy of the flat fields in the former band.
As a matter of fact, the uniform illumination required for flat fields 
could not be properly accomplished in focal reducer instruments due to 
scattered light and sky concentration. Moreover, this effect can be 
wavelength dependent (Andersen, Freyhammer, \& Storm 1995).     
In particular, we find that the I magnitudes in the two CCD regions 
located at $x>1300$ and $y<600$ pixels are less accurate, and 
indeed in these regions the scatter between current and Stetson's 
magnitudes is larger.   
 
As a further and independent test of the intrinsic accuracy of current 
calibration we compared our (V,V-I) CMD with 
the CMD recently provided by Rosenberg et al. (2000\footnote{Data 
available at the following Web site {\tt http://menhir.pd.astro.it}.}), 
and our (V, B-V) CMD with the CMD provided by CO97. The top panel of 
Fig. 2 shows that our data are in very good agreement with the 
Rosenberg et al. data, and indeed the main branches of the CMD nicely 
overlap. This evidence is further supported by the fact that both the 
magnitude ($V_{TO}=18.2$ mag) and the color ($(V-I)_{TO}=0.905$) of 
the Turn Off estimated by Rosenberg et al. are, within current 
uncertainties, in remarkable agreement with our evaluations. 

As far as the BV data is concerned, we found that the mean loci provided 
by CO97 seem slightly bluer than our diagram, and therefore we decided to 
check this discrepancy on a star by star basis. 
The bottom panel of Fig. 2 shows the difference in magnitude (open 
circles) and color (filled circles) between our and CO97's HB and RGB stars. 
Data plotted in this panel clearly show a color off-set that ranges 
from $\sim 0.02$ mag for red objects to $\sim 0.05$ mag for blue objects. 
On the other hand, the difference in the V magnitude steadily increases 
from blue to red objects. We suggest that the difference in the V band 
could be due to the procedure they adopted to calibrate the bright end  
of the CMD. In fact, their Danish data set ($V<16$) was calibrated  
with the NTT data set, i.e. with photometric data with  $16<V<18$ 
and $0.5<(B-V)<1$. As a consequence, their color term in the 
V magnitude calibration was extrapolated and possibly underestimated. 
This working hypothesis is supported by the fact that the difference in 
this color range is vanishing. Moreover, our determination of the 
distance modulus agrees quite well with their evaluation (see \S 6).

\section{Metal abundance}

The metal content is a key parameter to assess on a quantitative basis the 
properties of stellar populations. The mean metallicity of NGC~3201 has 
been a matter of concern for a long time. In fact, metal 
abundances based on integrated properties (Zinn 1980; Zinn \& West 1984, 
hereinafter ZW84), on low-dispersion spectra, and on the RGB slope in 
the (V,V-I) CMD (Da Costa, Frogel, \& Cohen 1981) do suggest for this 
cluster a mean metallicity 
similar to M~3 and NGC~6752, i.e. $[Fe/H]\approx -1.61$. However, 
Smith \& Manduca (1983) found -on the basis of the $\Delta S$ method 
applied to nine RR Lyrae- that the mean metallicity is $-1.34\pm0.15$. 
They also found no significant variation in the metal content among 
the RR Lyrae in the sample. On the other hand, Carretta \& Gratton (1997, 
hereinafter CG97), by re-analyzing high-dispersion CCD spectra (3 stars)
and on the basis of updated atmosphere models (Kurucz 1992), found a 
mean metallicity of $[Fe/H]=-1.23\pm0.09$ that is higher than previous 
ones. A slightly lower metal abundance was estimated by Carney (1996) 
$[Fe/H]=-1.34$ who also provided an $\alpha$-element overabundance of
$[\alpha/Fe] = 0.26$.  

The discrepancy between different empirical estimates was not solved by 
the extensive and homogeneous spectroscopic investigation, based on the 
equivalent width of the CaII triplet, provided by Rutledge et al. (1997). 
In fact, they found that the metallicity of NGC~3201 ranges from 
$[Fe/H]=-1.24$ to $[Fe/H]= -1.53$ according to the metallicity scales of 
CG97 and ZW84 respectively. The empirical scenario was further complicated 
by the fact that a spread in the metal content has also been suggested  
(Da Costa et al. 1981). This hypothesis was somehow supported by  
accurate spectroscopic measurements by Gonzales \& Wallerstein (1998, 
hereinafter GW98). They found that 
metallicity estimates for 17 cluster RG stars range from -1.17 to 
-1.68, and present variations that are roughly a factor of 4 larger 
than the typical uncertainty on individual [Fe/H] measurements. 
This notwithstanding, GW98 suggested that an intrinsic spread in the 
iron abundance among the RG stars of NGC~3201 is unlikely, and 
provided a mean metal content of -1.42.
According to this metal content, to the $\alpha$-element enhancement
estimated by GW98, $[\alpha/Fe]\simeq 0.40$, and by adopting 
the Salaris, Chieffi, \& Straniero (1993) relation, we estimate 
that the global metallicity\footnote{The global metallicity is a 
parameter which accounts for both iron and $\alpha$-element abundances 
(Carney 1996; Vandenberg 2000).} for this cluster is [M/H]=-1.13. 
In the following we will adopt this mean global metallicity and a 
mean iron abundance of [Fe/H]=-1.42.

\section{The RR Lyrae variables}

On the basis of BV, and I-band data we identified 64 of the 96 variables  
listed by Sawyer-Hogg (1973; hereinafter SH73), and by Samus et al. (1996). 
We confirm the non variability for V33, V70, V74, V75, V81, and V82 
quoted by SH73. Seven of the RR Lyrae we identified are located in a CCD 
region where the quality of the photometry is not accurate. 
Moreover, 2 out of the 8 new suspected variables 
identified by Lee (1977) and by Welch \& Stetson (1993), namely 2710 and 
1405, according to the photometric list by Lee (1977) have been confirmed 
as "true" variables. According to current photometry the suspected 
variables  -3516, 4702, 2517, 2403- do not show a strong evidence of 
variability, while for 1113 and 1103 we cannot reach any firm 
conclusion, since the quality of the photometry is poor.  

We performed a detailed {\em ad hoc} search aimed at detecting new 
RR Lyrae variables, but we did not find new variables and/or new candidates. 
Therefore, according to previous findings the number of RR Lyrae and  
suspected RR Lyrae present in NGC~3201 decreases from 104 to 94. 
Note that SH73 did not detect any variability for V79, a star located 
close to the tip of the RGB, whereas we found over the three nights a 
clear variation in the luminosity (see Appendix A for more details). 

Our data were collected over three consecutive nights, and therefore 
the light curves of RR Lyrae variables present a good phase coverage 
over the full pulsation cycle.   
The same outcome does not apply to RR Lyrae characterized by pulsation 
periods close to 0.5 days, since they present a poor phase coverage
close to maximum/minimum luminosity.   
Fig. 3 shows the atlas of BVI light-curves for the entire sample. 
The old variables were called with the name listed by SH73, while for 
the new ones we adopted the numbering introduced by Lee (1977).
Fig. 4 shows B and V light curves for the RR Lyrae with a poor phase
coverage close to maximum/minimum luminosity. To estimate the mean 
magnitudes and the amplitudes of these nine objects, we adopted the 
empirical light curve template provided by Layden (1998). The solid 
lines in Fig. 4 display the B and V template curves. Note that Layden 
only provided V band templates, but Borissova et al. (2001) found that 
these light curves, after a proper scaling, can also be adopted 
to derive the luminosity amplitudes in the B band. Data plotted in 
Fig. 4 show a fair agreement between the template light curves and 
available empirical data. However, for four (V18, V20, V38, V50) out 
of the nine variables the B band data present a systematic difference   
with the template close to the phase of minimum luminosity. This means 
that both mean magnitudes, colors, and amplitudes of these variables 
should be cautiously treated, since they are affected by larger 
errors.  

As far as the light curves are concerned, the photographic data set 
collected by C84 is still the most comprehensive and detailed 
investigation of RR Lyrae in NGC~3201. Fig. 5 shows the difference 
in the mean V (top panel), and B (bottom panel) magnitudes between 
current and C84 estimates.
Data plotted in this figure clearly show a very good agreement 
in the V magnitude, whereas the B magnitudes by C84 are 
systematically brighter by approximately 0.05-0.075 mag when 
compared with current estimates. However, such a difference is 
mainly due to a difference in the zero-point of the photometric 
standards (Lee 1977) adopted by C84. In fact, a detailed comparison 
between current CCD magnitudes for a dozen of Lee's photographic 
standards disclose the same shift in the B magnitude. 
 
\subsection{Amplitude modulation}

During the last few years several empirical (Kovacs 1995; Nagy 1998; 
Bragaglia et al. 2000; Smith et al. 1999) and theoretical 
(Shibahashi 2000; van Hoolst 2000) investigations have been focused 
on amplitude modulation among field RR Lyrae stars. In a series of papers 
Szeidl (1976, 1988, and references therein) showed that approximately 
one third of RR Lyrae show 
periodic or quasiperiodic variations in the luminosity amplitude. The time 
scale of the secondary modulation typically ranges from 11 days (AH Cam, 
Smith et al. 1994) to 530 days (RS Boo, Nagy 1998). The occurrence of this 
phenomenon, called {\em Blazhko effect} (Blazhko 1907), among RR Lyrae 
variables in GGCs (M~3, M~5, M~15, $\omega$ Cen) and in dwarf galaxies 
(Draco) was soundly confirmed by Smith (1981) who also found a similar 
frequency, i.e. 25-30\%. In this context it is worth mentioning that  
Corwin \& Carney (2001) recently provided very accurate BV CCD photometry 
for more than 200 RR Lyrae stars in M~3. The previous authors collected 
time series data over a time interval of 5 years, and therefore they 
unambiguously identified RR Lyrae variables that present amplitude 
modulations. The new data confirm the finding obtained by 
Smith, and indeed 47 out of 158 $RR_{ab}$ variables present the 
{\em Blazhko effect}, i.e. 30\% of the sample. Moreover,  
a recent detailed Fourier analysis based on the MACHO database does suggest 
that some fundamental Blazhko RR Lyrae present both amplitude and phase 
modulation, as well as the occurrence of the {\em Blazhko effect} among 
first overtone RR Lyrae (Kurtz et al. 2000).  

As a consequence, we decided to investigate the occurrence of such a 
phenomenon in our RR Lyrae sample. Fig. 6 shows the difference in the 
luminosity amplitude between current estimates and the luminosity 
amplitudes provided by C84. Data plotted in this figure disclose 
several interesting features: 
{\em i)} a fraction of 25-30\% among fundamental variables presents 
amplitude modulations; {\em ii)} one (V48) out of the three $RR_c$ variables 
in our sample shows amplitude modulation. This variable is the first 
candidate Blazhko $RR_c$ star in a GGC. 
The plausibility of these findings rely on the fact that all the RR Lyrae 
with amplitude modulations in the B band also show the modulation 
in the V band. Note that as a conservative estimate of the 
systematic difference between current photometry and the photographic 
photometry performed by C84, we only selected variables that show 
a difference in the luminosity amplitude larger than 0.1 and 
0.15 mag in the V and in the B respectively (dashed lines). 
The referee suggested to compare current mean magnitudes with the 
mean magnitudes provided by C84 to single out whether some of the candidate
Blazhko RR Lyrae are blends in one of the two data sets. Interestingly
enough, we found that the mean V magnitudes provided by C84 for variables
V31 and V36 are $\approx$0.12 and $\approx$ 0.20 mag brighter than current
ones (see Appendix A for more details).
This means that they could be blends in the former photometry. Oddly
enough, current mean V magnitude of variable V58 is 0.15 mag brighter
than the mean magnitudes provided by C84. The reason for this discrepancy
is not clear.  

On the basis of this finding, we decided to extend our analysis by 
covering a longer baseline, and, in particular, to compare current 
B amplitudes with the B amplitudes estimated by SH73. 
Fig. 7 shows the difference between current and SH73 mean B
magnitudes\footnote{Note that SH73 did not supply the mean B magnitudes,
therefore to estimate the difference we adopted the mean between minimum
and maximum.}. Data plotted in this figure display that mean
B magnitudes by SH73 are systematically brighter and the difference
ranges from a few hundredths up to $\approx$ 0.4 mag. According to
this evidence we will consider candidate Blazhko RR Lyrae stars the
objects that present a difference between current and SH73 B amplitudes
that is larger than 0.45 mag. The top panel of Fig. 8 
shows that approximately 30\% of RR Lyrae variables present 
amplitude modulations larger than 0.45 mag. The same outcome 
applies to V48 ($RR_c$ variable). It is noteworthy that data 
plotted in Fig. 8 seem to suggest that the difference in the 
B amplitude is mainly due to a difference in $B_{min}$ (middle 
panel) rather than in $B_{max}$ (bottom panel). The systematic 
difference in the two data sets do not allow us to constrain this 
effect on a more quantitative basis.  

We also note that approximately 50\% of RR Lyrae that show 
amplitude modulation in the SH73 sample have also been detected 
in the C84 sample, namely V6, V26, V34, V48, and V51.  
As a whole, the comparison of photometric data spanning a time 
interval of roughly 25 yr strongly support the evidence that 
a fraction of 25-30\% of RR Lyrae in NGC~3201 present amplitude 
modulation ({\em Blazhko effect}). Unfortunately, current data 
do not allow us to assess on a quantitative basis the secondary 
period. A comprehensive analysis based on new data as well as 
on old photographic data will be presented in a forthcoming 
paper (Bono et al. 2002, in preparation).   

In this context, we also mention that the comparison between 
predicted and empirical amplitudes among cluster variables 
should be handled with care. In fact, data plotted in Fig. 9 
clearly show that in the Bailey diagram both distribution 
and intrinsic scatter are somehow affected by the number of 
RR Lyrae that show the {\em Blazhko effect}. Note that current 
variations in the luminosity amplitude are indeed a lower limit 
to the "true" amplitude modulation. As a matter of fact, extensive 
data for field and cluster (M~3) RR Lyrae do suggest that the 
luminosity variation is, on average, of the order of half magnitude 
(see Fig. 3 in Szeidl 1988).  

Table 1 lists from left to right for each variable in our sample:  
1) the variable name, 2) the RR Lyrae type, 3) the period, 4) the 
reference for the period. Moreover, from column 5) to 7) we give 
the magnitude-weighted BVI mean magnitudes, while from column 8) to 10) 
the intensity-weighted BVI mean magnitudes. The last three columns 
-11 to 14- list the BVI luminosity amplitudes. The mean I magnitudes 
of RR Lyrae with noisy light curves, have not been included. However, 
a photometric scatter slightly larger than the typical value (0.01-0.02 mag) 
can also be found among BV light curves of variables located close 
to the edges of the CCD camera.


\section{Reddening evaluations}

\subsection{RR Lyrae (B-V) colors}

Accurate evaluations of the reddening across the central region of 
NGC~3201 are mandatory to supply robust estimates of intrinsic mean 
magnitudes and colors of RR Lyrae stars. However, precise reddening 
estimates are difficult and the problem for NGC~3201 is even more 
complicated, since this cluster presents patchy absorptions, i.e. 
differential reddening, across the main body of the cluster
(Zinn 1980; Alcaino \& Liller 1981; Da Costa et al. 1981); 
C84; CO97; Layden et al. 2002, hereinafter L02). 
As a matter of fact, current color excess estimates range 
from E(B-V)=0.21 to E(B-V)=0.29 according to mean RR Lyrae colors
(C84). On the basis of spectroscopic data for a sizable sample 
of red giants GW98 found E(B-V) values ranging from 0.21 to 0.31.  
The same outcome applies to mean reddening values, and indeed 
current values present a spread ranging from $0.21\pm0.03$ (C84), 
$0.22\pm0.03$ (CO97) to 0.28 (Harris 1976). 

A comprehensive analysis of the extinction map across NGC~3201 has 
been recently provided by vBM01 by adopting high-quality and deep 
V and I band data. In particular, 
they found a differential reddening of the order of 0.2 mag and a 
reasonable agreement on a large scale with the dust infrared maps 
provided by Schlegel, Finkbeiner, \& Davis (1998, hereinafter SFD98). 
At the same time,
they also confirmed the evidence originally brought out by Arce \& 
Goodman (1999) that infrared emissions maps overestimate reddening 
in high extinction regions ($E(B-V)> 0.15$). However, the zero point 
of the reddening scale found by vBM01 (E(V-I)=0.15) is approximately 
1.5 $\sigma$ smaller than mean values available in the literature.  
As a consequence, we decide to estimate the mean value of the color 
excess as well as of its variation across the cluster using the 
pulsation properties of our RR Lyrae sample.  

In their seminal investigations, Preston (1964) and Sturch (1966) 
calibrated an empirical relation for 
$RR_{ab}$ variables that supplies the color excess E(B-V) as a 
function of the (B-V) color at the phase of luminosity minimum, 
the period, and the metallicity. 
According to Walker (1990, hereinafter W90), who improved this empirical 
relation, we adopt:  $E(B-V) = (B-V)_{min} - 0.24P - 0.056*[Fe/H]-0.336$
where $(B-V)_{min}$ is the mean color over the pulsation phases 
$0.5\le \phi \le 0.8$, [Fe/H] is the metal content in the ZW84 
metallicity scale, and P the fundamental period (days).
On the basis of $(B-V)_{min}$ colors listed in column 3 of Table 2 
and by adopting the cluster iron abundance given by ZW84, 
$[Fe/H]=-1.61$, we found that the mean value of the 
reddening is $<E(B-V)> = 0.34\pm0.03$ and the individual values 
(column 4 in Table 2) range from 0.29 (V36) to 0.40 (V13). 
Individual estimates do suggest that across the cluster region 
covered by RR Lyrae in our sample the reddening changes  by 
approximately 0.1 mag. This finding supports the results 
obtained by GW98 on the basis of spectroscopic measurements of 17 
cluster RGs. On the other hand, the mean value estimated over the 
entire sample as a weighted mean should be cautiously treated. 
In fact, empirical evidence does suggest that reddening evaluations 
based on the Sturch's method are typically 0.03 mag larger than the 
determinations based on the slope of the RGB in the (V,V-I) CMD 
(Walker \& Nemec 1996). 

Therefore, we decided to provide new empirical evaluations 
of intrinsic (B-V) and (V-I) RR Lyrae colors by adopting two 
different routes. Since Schwarzschild (1940) it is well-known 
that RR Lyrae obey to a Period-Color (PC) relation. This 
evidence was further strengthened by empirical and theoretical 
investigations (Sandage et al. 1990a; Caputo \& De Santis 1992; 
Fernley 1993).
At the same time, theoretical predictions based on nonlinear, 
convective models suggest that the luminosity amplitude is 
strongly correlated with the effective temperature and presents  
a negligible dependence on stellar mass and metallicity 
(Bono et al. 1997a).
To derive robust empirical PC relations and Amplitude-Color (AC) 
relations we selected two GGCs, namely M~5 and IC~4499. These clusters 
present a sizable sample of RR Lyrae, have a metallicity similar to 
NGC~3201, and are not affected by differential reddening.    
According to the ZW84 metallicity scale the metallicity of these 
clusters is [Fe/H]=-1.5 (M~5) and [Fe/H]=-1.4 (IC~4499). Pulsation 
properties for RR Lyrae in M~5  were taken from Brocato, Castellani,
\& Ripepi (1996), Caputo at al. (1999), Storm, Carney, \& Beck (1991), 
and Reid (1996), while for RR Lyrae in IC~4499 from Walker \& Nemec (1996).  
 
However, the idea to derive intrinsic colors with either 
the PC or the AC relation was unsuccessful. In fact, the (B-V) and 
the (V-I) residuals clearly show a trend either with the luminosity 
amplitude or the mean color respectively. The reason why the 
empirical relations do not work is not clear. The failure of the 
PC relation in the optical bands might be due to the intrinsic width 
of the fundamental instability strip, as well as to the dependence 
on the metal abundance that causes an increase in the color scatter 
at fixed periods. The AC relation might be affected by the same 
drawback as well as by the scatter introduced by RR Lyrae affected  
by amplitude modulation.    
To overcome this thorny problem, we decided to derive, according to 
Caputo \& De Santis (1992, hereinafter CDS92), an empirical relation 
connecting the intrinsic color of RR Lyrae to the luminosity amplitude 
in the B band, and to the metallicity. To cover a wide metallicity range 
we selected field RR Lyrae stars observed by Lub (1977) and by 
Carney et al. (1992). As far as the Lub's sample is concerned we 
adopted the pulsation parameters collected by Sandage (1990b). 
Objects that present amplitude modulations have been neglected. 
We ended up with a sample of 78 RR Lyrae whose  metallicity ranges 
from [Fe/H]=-2.2 to [Fe/H]=0. We found that these objects do obey 
to a well-defined Amplitude-Color-Metallicity (ACZ) relation:

\[ 
(B-V)_0=0.448(\pm0.017) - 0.078(\pm0.006)A_B + 0.012(\pm0.004)[Fe/H] \;\;\;\;\; \sigma=0.016  
\]  

where the symbols have their usual meaning and units. Both the constant 
term and the coefficient of the luminosity amplitude 
are quite similar to the values found by CDS92 (see their relation 10). 
On the other hand, the coefficient of the metallicity term is almost a 
factor of two larger (0.012 vs 0.006) in the current relation than in 
the CDS92's relation. The difference seems to be due to the increase in 
the sample size, and to the inclusion of metal-rich  RR Lyrae 
(Carney et al. 1992).    
 
To check whether the current relation could be further improved, 
we correlated the intrinsic mean (B-V) color to period, luminosity 
amplitude, and metallicity. The linear regression over the same 
RR Lyrae sample supplies the following Period-Amplitude-Color-Metallicity 
(PACZ) relation:

\[ 
(B-V)_0=0.507(\pm0.014) - 0.052(\pm0.007)A_B + 0.223(\pm0.039)\log P + 
0.036(\pm0.005)[Fe/H] \;\;\; \sigma=0.014  
\]  
 
Note that such a relation relies on the assumption that the current 
sample of RR Lyrae stars is representative of the entire population. 
To validate the accuracy of these relations based on field RR Lyrae,  
we applied it to RR Lyrae in M~3, since this cluster is only marginally 
affected by reddening ($E(B-V)\approx0.01$, Dutra \& Bica 2000, 
hereinafter DB00). We selected, among the RR Lyrae variables (207) 
observed by Corwin \& Carney (2001) in M~3, the fundamental ones that 
are not affected by blends and present accurate estimates of both 
periods and B amplitudes. We ended up with a sample of 127 RR Lyrae. 
By assuming a metal content for M~3 of [Fe/H]=-1.46 (Kraft et al. 1995) 
and by applying the ACZ relation to this sub-sample, we found a mean 
cluster reddening of $<E(B-V)>=0.008\pm0.027$, while by adopting the 
PACZ relation we found $<E(B-V)>=0.007\pm0.024$. To avoid any spurious 
effect, if any, introduced by RR Lyrae that present the {\em Blazhko 
effect},   
we only selected "canonical" RR Lyrae (81). On the basis of the 
new sample, the ACZ and the PACZ relation give $<E(B-V)>=0.012\pm0.027$ 
and $<E(B-V)>=0.011\pm0.024$ respectively. 
The difference in the mean cluster reddenings based on the two different 
samples is negligible due to the marginal dependency of mean colors on  
amplitude modulations (see crosses in Fig. 10). 
The previous estimates are, within the errors, in 
very good agreement with the values given in the literature. 
The difference between ACZ and PACZ relation is negligible. 
However,  color estimates based on the latter one present a 
slightly smaller intrinsic scatter. 
 
As a further independent test we applied the previous relation 
to fundamental RR Lyrae not affected by amplitude modulation 
in several GGCs. We found that the mean reddenings as well as 
the dispersions are, within the uncertainties, in very good 
agreement (see data listed in column 5 of Table 3) with similar 
estimates available in the literature and based on stellar colors 
or on far-infrared dust emission (SFD98; DB00). 
On the basis of this evidence we estimated the individual reddenings,  
and in turn the intrinsic (B-V) colors of RR Lyrae in our sample 
(see data listed in columns 5 and 6 of Table 2). We found that the 
mean reddening is $<E(B-V)> = 0.30\pm0.03$ if we assume 
[Fe/H]=-1.42, while the single values range from 0.22 (V36) to 
0.35 (V45). Note that individual reddening estimates based 
on (B-V) colors are typically affected by an 
uncertainty of the order of 0.03 mag. The error budget includes the 
uncertainty affecting the photometric calibration 
($\sigma_{B,V}\approx 0.02$ mag), the intrinsic scatter of 
the PACZ relation ($\pm 0.01$ mag), as well as the error on the 
assumption that the reddening toward M~3 is E(B-V)=0.01. We did not 
account for the uncertainty in the mean metallicity, and indeed 
the mean reddening ranges from $<E(B-V)>=0.29\pm0.03$ to  
$<E(B-V)>=0.31\pm0.03$ if we assume [Fe/H]=-1.23 (CG97) or  
[Fe/H]=-1.61 (ZW84). 
Current mean reddening values, if we account for the entire error 
budget, support the mean reddening value obtained by adopting the 
SFD98 map. In fact, the mean reddening provided by this map over 
the same cluster region covered by our RR Lyrae sample is 
$<E(B-V)>\approx0.26\pm0.02$. This finding supports the absolute 
zero point of the SFD98 map.

\subsection{RR Lyrae (V-I) colors}

We collected photometric data in three different bands, and therefore we 
can constrain the occurrence of systematic errors, if any, in the reddening 
evaluations based on (B-V) colors. To estimate the intrinsic (V-I) colors 
we selected among field RR Lyrae stars the objects, as for (B-V) colors, 
for which accurate estimates of (V-I) color, metallicity, and reddening 
are available. We ended up 
with a sample of 18 $RR_{ab}$ stars (see Table 4). However, we realized 
that this sample is affected by a selection bias. In fact, all of them  
were originally chosen to perform the Baade-Wesselink analysis, 
and therefore 
they present large luminosity amplitudes and are typically located 
close to the blue edge. As a consequence, we decided to include two GGCs 
that host a sizable sample of RR Lyrae and accurate V and I photometry, 
namely IC~4499 (Walker \& Nemec 1996, $N_{RR}=35$) and NGC~6362 
(Walker 2001, private communication, $N_{RR}=14$). 
On the basis of these data, we found that 
they do obey to the following Period-Amplitude-Color (PAC) relation:    
 
\[ 
(V-I)_0= 0.65(\pm0.02) - 0.07(\pm0.01)A_V + 0.36(\pm0.06)\log P  
\;\;\;\;\; \sigma=0.02  
\]  

where the symbols have their usual meaning. This relation when compared 
with the previous one, presents a key difference:  
the coefficient of the metallicity term is vanishing, and therefore the 
linear regression was performed by neglecting this parameter. This 
effect is due to the fact that the (V-I) colors present a mild dependence 
on metallicity. However, we cannot firmly assess whether this is an 
intrinsic behavior of RR Lyrae stars. 
In fact, RR Lyrae listed in Table 4 together with RR Lyrae in
IC~4499 and NGC~6362 cover a wide metallicity range but only one object
is more metal-poor than [Fe/H]=-1.7. 

To estimate the accuracy of this relation it was applied to cluster RR Lyrae 
for which homogeneous estimates of mean (B-V) and (V-I) colors are available.  
Column 6 of Table 3 gives the mean reddening values and the standard 
deviations. As a whole, we found that reddening estimates based on 
(V-I) intrinsic colors are in good agreement with those based on (B-V) 
colors, and indeed the $<E(V-I)>$ values are, within the errors, 
approximately equal to $1.22\times<E(B-V)>$ (Cardelli et al. 1989; 
Bessell 1979). However, the mean reddening of NGC~1851 based on (V-I) 
colors is smaller than the reddening based on (B-V) colors. It has 
been recently claimed (Kovacs \& Walker 2001, hereinafter KW01) 
that the zero-point 
of the I-band photometry of this cluster could be affected by a 
systematic error. This notwithstanding the agreement between the 
two independent reddening estimates strengthens the plausibility 
and the accuracy of previous PACZ and PAC 
relations. Any further analysis concerning the difference between 
the two reddening scales is premature. In fact, the accuracy of 
the absolute zero-point of the PAC relations was not checked with 
a template cluster such as M~3, since current I band 
data could still be affected by uncertainties in the absolute 
zero-point calibration (Ferraro et al. 1997).   

However, individual reddening estimates, based on the PAC relation 
listed in Table 5, are in very good agreement with the evaluations 
based on the PACZ relation, and indeed the bulk of them attain values
roughly equal to $1.22\times <E(B-V)>$. Moreover, we find that the mean 
reddening is $<E(V-I)> = 0.36\pm0.05$, while the single values range 
from 0.28 (V36) to 0.45 (V45). Note that the reddenings based 
on (V-I) colors are affected by an uncertainty of the order of 0.04 mag. 
The error budget includes the error on the photometric calibration 
($\sigma_{I}\approx0.03$) and on the intrinsic scatter of the PAC 
relation ($\sigma_{I}\approx0.02$).    
This notwithstanding current results strongly support the finding 
obtained by vBM01 on the basis of TO star colors. In fact, we find 
that across the cluster, the E(V-I) values undergo differential 
changes up to $\sim0.17$ mag. This differential variation agree 
quite well with the reddening estimates by vBM01, and indeed across 
the same cluster region (see their Fig. 4) their reddening estimates 
(E(V-I)) range from 0.18 to 0.35 respectively.
On the other hand, our estimates do suggest a zero-point larger 
than evaluated by vBM01. As a matter of fact,  
we find that the mean reddening in the cluster region covered by our 
observations is $<E(V-I)>=0.36\pm0.05$, while the mean reddening 
based on the vBM01 map supplies $<E(V-I)>=0.25\pm0.07$. The reasons 
for this difference are not clear. A plausible reason could be the 
effect of hetero-chromatic extinction between TO and HB stars (Roberts 
\& Grebel 1995; Anthony-Twarog \& Twarog 2000). Note that the current 
reddening estimate agrees, within the errors, with the mean reddening 
obtained using the SFD98 map, and indeed we find 
$<E(V-I)>=1.22\times <E(B-V)>=0.32\pm0.02$. 

To constrain once again the intrinsic accuracy of current reddening 
estimates, we decided to perform a new test. According to a 
well-established result, the minimum-light color of $RR_{ab}$ 
stars present a mild dependence on metallicity (Lub 1979). This 
evidence has been further strengthened by Mateo et al. (1995, 
hereinafter M95) who found that the mean minimum-light (V-I) colors 
of a dozen of well-observed field RR Lyrae is roughly equal to 
$0.58\pm0.03$ mag. We estimated once again the individual reddenings, 
see column 5 in Table 5, and, interestingly enough, the new estimates 
are in good agreement 
with reddening evaluations based on the PAC relation. As a matter 
of fact, we found $<E(V-I)>=0.36\pm0.05$, while individual reddenings
range from 0.24 (V36) to 0.46 (V4).

\subsection{Fourier coefficients}

During the last few years several investigations have been devoted to 
the Fourier analysis of empirical light curves of RR Lyrae variables
(Kovacs \& Kanbur 1998).   
The underlying idea of these studies is to derive empirical relations  
based on Fourier coefficients that can be safely adopted to estimate 
physical parameters, such as absolute magnitude, intrinsic color, 
and metallicity of variable stars (Simon \& Clement 1993; Jurcsik 
\& Kovacs 1996; KW01). Therefore we followed this approach to supply 
an independent estimate of individual RR Lyrae reddenings. 
We performed at first the fit of the V-band light curves by means of 
sine Fourier series by adopting 15 components and estimated the 
typical Fourier coefficients. Columns 2 to 6 of Table 6 list the first 
Fourier amplitude, as well as amplitude ratios ($R_{21}$,$R_{31}$) and 
phase differences ($\phi_{21}$,$\phi_{31}$). Then we estimated the 
intrinsic mean $(B-V)_0$ and $(V-I)_0$ colors according to the empirical 
relations derived by KW01 and based on period as well as on $A_1$, and 
$A_3$ Fourier amplitudes (see their relations 6 and 9). The individual 
reddening 
evaluations based on these relations are given in columns 5 and 6 of 
Table 6. The new estimates soundly confirm previous findings, and 
indeed E(B-V) values range  from 0.24 (V36) to 0.35 (V13,V56), 
while the E(V-I) values range from 0.30 (V22,V36) to 0.44 (V56). 
Moreover, the mean values we find are:  
$<E(B-V)>=0.31\pm0.03$, and $<E(V-I)>=0.36\pm0.04$ respectively.  

Although, different methods based on mean (B-V) and (V-I) colors do 
supply similar reddening estimates we decided to compare individual 
values to investigate whether they present any systematic drift with 
intrinsic color. Fig. 10 shows the difference between current reddening  
evaluations based on the PACZ relation and the estimates based on the 
W90 relation (top), on the KW01 relation (middle), as well as on the 
ACZ relation 
(bottom). Data plotted in the top panel clearly show the systematic 
difference of $\approx 0.03$ between the two methods (see \S 5.1). 
On the other hand, data plotted in the middle and in the bottom panel 
clearly indicate that the difference between the PACZ, the KW01, and 
the ACZ relations is negligible. However, data plotted in the middle 
panel seem to suggest that the scatter of individual measurements 
increases toward the blue edge of the instability strip. 
The relations derived by KW01 do rely on well-observed cluster RR Lyrae 
and therefore this finding strengthens the accuracy of current relations 
and supplies and independent support to the KW01 relations.  

Fig. 11 shows the difference between current reddening determinations 
based on the PAC relation and the estimates based on the M95 (top) and 
on the KW01 (bottom) relation.
A glance at the data plotted in the top panel seems to suggest that 
the M95 relation slightly underestimates reddening evaluations when 
compared with reddenings based on the PAC relation. According to current 
data sample it is not clear whether this discrepancy increases  
when moving toward the blue edge. On the other hand, data 
plotted in the bottom panel disclose that reddenings based on the 
PAC relation and on the KW01 relation are in very good agreement, within 
the uncertainties. The evidence that close to the blue edge the scatter 
of individual measurements between current and KW01 relations increases 
also applies to the E(V-I) values.  
The variable V45 (open circle) presents a peculiar color excess,  
see Appendix A for more details.

\section{Comparison between theory and observations}

To supply accurate estimate of the mean RR Lyrae magnitudes and 
colors in our sample we performed a weighted mean between the E(B-V)  
determinations obtained by adopting both the PACZ and the KW01 
relation. The same approach was adopted to evaluate the mean E(V-I),  
but according to the PAC and to the KW01 relation. Table 7,
lists from left to right the identification, the dereddened V magnitude,
(B-V), and (V-I) colors, together with the color excesses, namely E(B-V) and 
E(V-I). The reddenings of $RR_{ab}$ stars whose light curve was fitted with 
the Layden's template are only based on the PACZ and the PAC relation. 
Magnitude and colors of $RR_c$ stars were dereddened by smoothing with a 
spline the reddening map derived using $RR_{ab}$ color excesses. 

According to previous individual evaluations, 
$<V_0(RR)>=(\sum_{i=1}^{nr} [V(RR) - 3.1\times E(B-V)])/nr$, 
we estimated the apparent Zero Age Horizontal Branch (ZAHB) luminosity 
by adopting the relation originally suggested by 
Carney et al. (1992) and revised by 
Cassisi \& Salaris (1997, hereinafter CS97), i.e. 
$V(ZAHB)= <V(RR)> + 0.04 [M/H] + 0.15$.       
The comparison between predicted ZAHB luminosities for [M/H]=-1.13  
(CS97) and empirical V magnitudes dereddened  
by adopting previous E(B-V) evaluations supplies a distance modulus 
for NGC~3201 of DM=$13.30\pm0.08$ (nr=40). Interestingly enough, 
if we adopt the visual ZAHB magnitudes dereddened by adopting the E(V-I) 
determinations, 
$<V_0(RR)>=(\sum_{i=1}^{nr} [V(RR) - 2.54\times E(V-I)])/nr$, 
we find DM=$13.35\pm0.09$ (nr=35).  
Fig. 12 shows the comparison in the $V_0-(B-V)_0$ and in the 
$V_0-(V-I)_0$ CMD between predicted ZAHB magnitudes (solid line) 
and observed HB stars. The dashed line marks the exhaustion of 
central He burning. Predicted luminosities and effective temperature 
were transformed into the observational plane by adopting the 
bolometric corrections and the color-temperature relations by 
Castelli, Gratton, \& Kurucz (1997).  
  
Magnitudes and colors of HB stars plotted in Fig. 12 were dereddened 
by adopting the same procedure adopted for $RR_c$ variables, and 
therefore they are the subsample of cluster HB stars located in 
the cluster region covered by current RR Lyrae sample. 
Data plotted in this figure clearly show that theoretical predictions 
for central He burning structures are, within 
current uncertainties on reddening and distance estimates in 
good agreement with observations. Stars located below the ZAHB 
with $0.4 \le (B-V) \le 0.5$ and $0.55 \le (V-I) \le 0.65$ colors 
should be field stars, and indeed a similar plume shows up just 
above the sub giant branch (see Fig. 15). 
A few RR Lyrae are also located below the ZAHB, but they present a
limited phase coverage or they are affected by amplitude modulation.
Note that in the latter case current B and V amplitudes could be
underestimated, and in turn the individual reddening corrections
(see \S 5.1 and 5.2). The region of hot HB stars presents a larger 
spread in the (V,V-I) than in the (V,B-V) diagram. This is due to 
the fact that magnitudes and colors based on E(V-I) determinations  
present larger errors than those based on E(B-V) ones (see the 
error bars in Fig. 12). 
 
Accurate distance moduli of globular clusters can be provided by 
adopting the method suggested by Caputo (1997). 
The match in the $M_V-\log P$ plane between predicted FO blue 
edge and the observed blue limit of $RR_c$ variables does supply 
an independent and robust estimate of the distance modulus 
(Caputo et al. 2000, and references therein). Unfortunately, the 
application of this method to NGC~3201 is hampered by the small 
number of $RR_c$ variables (5), as well as by the fact that 
we lack individual reddening estimates for these objects. 
However, to supply an independent check on the accuracy of current 
distance estimates Fig. 13 shows the comparison in the (V,B-V) and 
in the (V,V-I) plane the comparison between 
predicted instability edges for $RR_{ab}$ (solid lines) and $RR_c$ 
(dashed lines) pulsators (Bono et al. 2001, and references therein) 
and observed 
RR Lyrae stars in NGC~3201. Predicted edges were plotted by assuming a 
distance of DM=13.32 and rely on full amplitude, nonlinear, convective 
models constructed by adopting the same stellar mass 
($M/M_\odot= 0.70$) and 
chemical composition (Y=0.24, Z=0.0004) and a 
wide range of stellar luminosities. Luminosities and effective 
temperatures were transformed into the observational plane by 
adopting the transformations by Castelli et al. (1997).  

Note that the distribution of variable stars inside the instability 
strip is in good agreement with theoretical predictions. In fact,  
the red (cooler) edge properly marks the  
transition between $RR_c$ and $RR_{ab}$ variables. On the other 
hand, the RR Lyrae in the (V,V-I) plane appear slightly redder 
thus supporting the evidence that we are slightly underestimating 
the reddening in this plane. Intrinsic colors of $RR_c$ variables 
should be cautiously treated, since their reddenings are based on 
the $RR_{ab}$ reddening map and therefore affected by uncertainties 
of the order of E(V-I)$\approx 0.04$. 
The metallicity adopted for the instability edges ([Fe/H]=-1.65) is 
more metal-poor than adopted for HB models ([Fe/H]=-1.42, i.e. 
[M/H]=-1.13). The 
comparison between empirical data and predicted edges for Z=0.001 
shows the same agreement. The main difference is that a good fraction 
of $RR_{ab}$ variables moves into the "OR" region. Unfortunately, 
the sample of $RR_c$ variables is too small to draw any firm 
conclusion concerning the topology of the instability strip and 
the mean metallicity. 
 
The luminosity amplitudes are robust observables to constrain the 
physical parameters governing the pulsation properties of radial 
variables. In fact, they are not affected by empirical uncertainties 
in the distance modulus, and marginally affected by errors in reddening 
evaluations. 
Fig. 14 displays the comparison in the Bailey diagram between current 
sample of RR Lyrae variables with RR Lyrae in M~5 (open circles) and 
in IC~4499 (open squares). We selected these two clusters because 
they have almost the same metallicity, and host a sizable sample 
of RR Lyrae. Data plotted in this figure indicate that RR Lyrae in 
these three clusters do show similar distributions in the Bailey 
diagram. In particular $RR_{ab}$ variables do cover the same period 
range and attain similar luminosity amplitudes when moving from the 
blue to the red edge of the instability strip. The number of $RR_c$ 
variables in NGC~3201 is too small to draw any conclusion concerning 
the occurrence of a systematic difference in the period distribution. 

Data plotted in Fig. 14 clearly support 
the finding originally brought out by Smith (1981): {\em irregular 
variability is more frequent among fundamental pulsators with 
periods shorter than 0.65 days}. As a matter of fact, the spread 
in the luminosity amplitudes becomes substantially smaller for 
$\log P \ge -0.2$.  
Together with empirical data Fig. 14 shows the comparison with 
predicted B,V, and I-band amplitudes. Theoretical observables 
are based on the same models we adopted in Fig. 13. Predicted 
amplitudes for $\log L/L_\odot=1.61$ and 1.72 bracket empirical 
data and suggest a luminosity for RR Lyrae in NGC~3201 of the order 
of $\log L/L_\odot=1.65$. This pulsational estimate agrees, within 
current uncertainties on chemical composition, with the luminosities 
predicted by HB models when moving from [Fe/H]=-1.25 to 
[Fe/H]=-1.65, i.e. $1.63 \le \log L/L_\odot \le 1.68$.   
Finally, we note that empirical amplitudes for $RR_{ab}$ stars 
do show, at fixed period, a large scatter inside the instability 
strip. However, the RR Lyrae in NGC~3201 that show amplitude 
modulation are located at the bottom of this distribution. 
This finding supports the evidence that the observed scatter could 
be due to RR Lyrae that undergo amplitude modulations.

\section{The Color-Magnitude diagram}

The top panels of Fig. 15 show the (V,B-V) and (V,V-I) diagrams. 
The effect of the differential reddening 
is quite evident in the (V,V-I) diagram, and indeed in this plane the 
stars along the RGB split over two distinct sequences. This 
notwithstanding the CMDs show a well-defined Main Sequence (MS) at least 
down to V$\approx$20.5, well-populated post-MS phases together with a 
sequence of Blue Stragglers approaching hot HB stars. Data plotted in 
this figure are the average of more than 70 measurements in each band. 
We only plotted stars with photometric errors in the three bands 
smaller than 0.02. The bottom panels of Fig. 15 display the extinction 
corrected $V_0$ magnitudes vs $(B-V)_0$ and $(V-I)_0$ obtained 
by adopting the reddening maps based on $RR_{ab}$ stars.
The number of stars in the bottom panels is smaller than in the 
top ones, since we only plotted stars located inside current reddening 
maps. The resolution of these maps is of the order of 1 arcmin. 
The plausibility of the reddening correction is supported by the 
narrowing of the main evolutionary phases. 

The CMD of NGC~3201 presents several similarities with the CMD of 
M~3, and indeed the fiducial lines derived by Ferraro et al. (1997) 
for this cluster overlap to dereddened data for NGC~3201 (see bottom 
left panel of Fig. 15). The main discrepancy between the two cluster 
data sets is among extreme HB stars. This mismatch could be due to 
the calibration of the photometric zero-point (see Ferraro et al. 1997 
for further details). At the same time, we also note that the (B-V) color 
of our standard stars do not cover extreme HB stars. This notwithstanding, 
data plotted in this figure show a very good agreement between the RGB 
mean loci of M~3 and individual RGB stars in NGC~3201.  
This evidence supports the zero-point of current reddening scale as well 
as individual reddenings, since M~3 is only marginally affected by 
reddening (E(B-V)$\approx 0.01$). A further interesting similarity between 
M~3 and NGC~3201 is a sizable sample of Blue Stragglers (CO97). 
The properties of these stars will be addressed in forthcoming paper. 
The same outcome applies to the dereddened (V,V-I) CMD, and indeed 
data plotted in the bottom right panel show a very good agreement 
between the mean ridge line provided by Johnson \& Bolte (1998) and 
current photometry. Note that the agreement applies not only to 
TO and RGB regions but also to hot and cool HB stars.  

A thorough discussion concerning the evolutionary properties  
of this cluster will be addressed in a forthcoming paper. In 
this context, we briefly discuss the RGB bump. This interesting 
evolutionary feature appears as a peak in the differential 
Luminosity Function (LF) and 
as a change in the slope of the cumulative LF. From an evolutionary 
point of view, the presence of such a bump is due to the fact that during
the RGB evolution the H-burning shell crosses the chemical discontinuity 
left over by the convective envelope soon after the first dredge-up at 
the base of the RGB. We located the RGB bump in NGC~3201 using both 
the differential and the cumulative LFs and we find 
$V_{bump}=14.55\pm0.05$ mag. The comparison between theory and observations 
concerning the location of the bump along the RGB is a crucial test to  
assess the accuracy of current evolutionary models. Dating back to 
Fusi Pecci et al. (1990) it became clear that the $\Delta{V}_{HB}^{bump}$, 
i.e. the difference in magnitude between the RGB bump and the HB stars 
located inside the RR Lyrae instability strip, is the key parameter to 
compare theory and observations.  

To estimate this parameter we adopted current RR Lyrae sample and we 
find a mean magnitude of $<V_{RR}>=14.75\pm0.07$. This magnitude was 
scaled to the ZAHB luminosity according to the correction suggested 
by CS97 and eventually we find $V_{ZAHB}=14.85\pm0.07$ mag. 
Therefore, the $\Delta{V}_{HB}^{bump}$ for NGC~3201 is equal to 
$-0.30\pm0.09$ mag. Fig. 16 shows the comparison between this value 
and theoretical predictions (CS97). Note that for NGC~3201 we adopted 
the mean global metallicity based on the spectroscopic measurements of 
iron and $\alpha$-elements given by GW98, and the relation between global 
metallicity, iron abundance, and $\alpha$-element enhancements  
provided by Salaris et al. (1993). 
For the aim of the comparison, the same figure also shows empirical  
estimates of $\Delta{V}_{HB}^{bump}$ for a selected sample of GGCs 
for which accurate spectroscopic measurements of iron and 
$\alpha$-element enhancement are available (see labeled names). 
Theoretical predictions rely on evolutionary models at fixed initial 
He content (Y=0.23, Zoccali et al. 2000) that cover a wide range of 
progenitor masses ($M/M_\odot=0.8-1.0$), and global metallicities 
($-2.3\le [M/H]\le -0.5$). Data plotted in this figure show that 
theory and observations are in fine agreement. 
The empirical value of $\Delta{V}_{HB}^{bump}$ for NGC~3201 seems 
slightly smaller than predicted by theoretical models. This finding 
taken at face value could suggest that the mean global metallicity 
of this cluster is slightly more metal-poor than currently estimated.

\section{Summary and conclusions}

We present BVI time series data of NGC~3201 collected over 
three consecutive nights that cover a cluster region of approximately 
13 arcmin$^2$ around the center. Current data allowed us 
to identify 72 out of the 104 variables or suspected variables 
originally detected by SH73, Lee (1977), and by Welch \& Stetson (1993). 
According to current data we confirm the non variability for ten of 
them and strongly support the evidence brought out by GW98 that V79 
is a red variable located close to the tip of the RGB. The light curves 
of RR Lyrae in our sample (53) present a very good phase coverage. 
The only exception to this rule are 9 RR Lyrae with periods close 
to 0.5 days that present a poor phase coverage close to the phases 
of maximum/minimum luminosity. 

The comparison between current data and photographic data collected 
by SH73 and C84 support the evidence that approximately the 30\% 
of $RR_{ab}$ variables present amplitude modulation ({\em Blazhko 
effect}). This result supports the findings originally brought out 
by Szeidl (1976) and Smith (1981) for field and cluster RR Lyrae 
stars. At the same time, we confirm that $RR_{ab}$ that show 
the {\em Blazhko effect} present periods shorter than 0.65 days
(Smith 1981). We also find that one (V48) out of the three $RR_c$ 
variables is strongly suspected to be affected by amplitude modulation.  
If new data confirm this evidence, this object would be the first 
cluster $RR_c$ variable that shows such a phenomenon.  

To overcome the thorny problem of differential reddening across 
the cluster we derived new empirical relations connecting the 
intrinsic (B-V) and (V-I) colors of RR Lyrae stars with pulsation 
parameters. By adopting a large sample of field 
RR Lyrae (78) for which are available B amplitude, (B-V) color, 
metallicity, and present low reddening or accurate reddening 
estimate we find that they do obey to an ACZ and to a PACZ 
relation. The key features of these relations are the following: 
{\em i)} they depend on stellar parameters that are not affected 
by reddening. {\em ii)} They supply accurate estimates of intrinsic 
(B-V) colors across the fundamental instability region and cover a wide 
metallicity range. This means that they can be used to estimate the 
reddening of halo and bulge RR Lyrae stars.   
{\em iii)} They have been derived by neglecting the RR Lyrae 
that present or are suspected to be affected by amplitude 
modulation. 
To validate the zero-point of the reddening scale  
we applied the new relations to the large sample of RR Lyrae 
stars (207) in M~3 recently collected by Corwin \& Carney (2001). 
We selected this RR Lyrae sample, since the reddening toward 
this cluster is vanishing (E(B-V)$\approx0.01$). We find that 
both the ACZ and the PACZ relations account for the mean 
cluster reddening, but the intrinsic scatter of the latter 
one is somehow smaller. Moreover and even more importantly, 
we find that the difference in the mean cluster reddening based 
on an RR Lyrae sample that includes or neglects Blazhko RR Lyrae 
is marginal. This finding suggests that previous relations can 
be safely adopted to estimate the reddening of well-sampled 
Blazhko RR Lyrae.  

We adopted the same approach to derive a relation that allow us 
to derive individual reddening on the basis of the (V-I) colors.  
We selected a sample of 18 field plus 49 cluster (IC~4499, 
NGC~6362) $RR_{ab}$ variables for which are available V amplitude, 
(V-I) color, metallicity, and accurate reddening estimates. 
Interestingly enough, we find that they do obey to a PAC relation, 
i.e. the metallicity term is vanishing. This means that this relation 
can be used to estimate the reddening on the basis of period 
and mean (V-I) color. Unfortunately, we could not validate the 
zero-point of this relation, since I-band data are not available 
for RR Lyrae in M~3, and we are not aware of GGCs with a large 
sample of RR Lyrae, a vanishing reddening, and accurate (V-I) colors. 
However, the comparison between the mean reddenings  
based on the PAC and on the PACZ relation for clusters for which are 
available BVI data shows a good agreement, within the uncertainties.  

We applied the PACZ relation to fundamental RR Lyrae in our sample 
and by assuming a mean cluster metallicity of [Fe/H]=-1.42 we find 
$<E(B-V)>=0.30\pm0.03$, while the individual reddening estimates
range from 0.22 (V36) to 0.35 (V45). The intrinsic scatter we find 
($\Delta E(B-V)\approx0.15$) is slightly larger than estimated by 
GW98 on the basis of 17 RGB stars. However, it is worth noting, that 
current estimate agrees with the mean value obtained by 
adopting the dust infrared map by SFD98. As a matter of fact, 
on the same cluster region we find a mean reddening of 
$<E(B-V)>=0.26\pm0.02$. Note that the angular resolution of the 
SFD98 map is approximately 6 arcmin, while our map has a resolution 
of the order of 1 arcmin.   

The same outcomes apply to the PAC relation, and indeed we find 
a mean cluster reddening of $<E(V-I)>=0.36\pm0.05$, while the 
individual reddening estimates range from 0.28 (V36) to 0.45 (V45).   
The spread between individual evaluations support the results by 
vBM01 who found that the differential reddening across the 
cluster changes by approximately 0.2 mag. On the other hand, 
current mean cluster reddening is larger than the mean reddening 
obtained using the reddening map by vBM01 on the same cluster 
region ($<E(V-I)>=0.25\pm0.07$). Although, our evaluation based 
on (V-I) colors is internally consistent with the mean reddening 
based on the (B-V) colors, $<E(V-I)>=1.22 \times <E(B-V)> = 0.37\pm0.04$, 
and in turn with the SFD98 map.   
   
We performed a detailed comparison with similar relations 
available in the literature and we found that current relations 
are in very good agreement with the reddening estimates based 
on the relations derived by KW01. In fact, the KW01 relations 
supply mean cluster reddenings of ($<E(B-V)>=0.31\pm0.03$) 
and of ($<E(V-I)>=0.36\pm0.04$). These relations have been 
calibrated using cluster RR Lyrae for which are available BV 
and I band data and rely on Fourier amplitudes and periods. 
The main advantage of current relations is that they can also 
be applied to RR Lyrae with limited phase coverage.  
However, the ACZ and the PACZ relations 
require the knowledge of the metallicity, while the KW01 
relations and the PAC relation only rely on periods and 
luminosity amplitudes.   

To supply a detailed comparison between theory and observations 
we dereddened the magnitudes of the stars located inside current  
reddening maps. The comparison between predicted 
ZAHB luminosities for He burning structures located inside 
the instability strip and observed RR Lyrae stars provides  
a distance modulus of $13.30\pm0.08$ if we adopt the V magnitudes
dereddened using the E(B-V) map. Interestingly enough, 
we find that the distance modulus is $13.35\pm0.09$ if the 
V magnitudes are dereddened using the E(V-I) map, and therefore the 
weighted true distance modulus is $13.32\pm0.06$. This estimate is 
in good agreement with values available in the literature and based 
on different distance indicators (see CO97 for a detailed discussion). 
Although, the uncertainty affecting current distance determination is 
smaller than similar estimates. The main difference is due to the fact 
that we are adopting individual reddenings and not a mean cluster 
reddening. However, previous errors do not account for current 
uncertainties on predicted ZAHB luminosities ($\approx0.1$ mag, see 
e.g., Cassisi et al. 1999, Bono, Castellani, \& Marconi 2000) 
as well as on mean cluster metallicity. An uncertainty of 0.2 dex 
in the mean metallicity implies an uncertainty on the distance modulus 
given by predicted ZAHB luminosities of the order of 0.05 mag. 
It goes without saying that the K-band Period-Luminosity-Metallicity 
relation of RR Lyrae should overcome these problems, since it 
presents a mild dependence on cluster metallicity as well as on 
individual reddenings (Bono et al. 2001). 

Current RR Lyrae sample covers a cluster region that is 
approximately one third of the cluster tidal radius 
($r_t = 28.5$ arcmin, 
Harris 1996\footnote{{\tt http://physun.mcmaster.ca/~harris}.}). 
New multiband time series data collected with a wide field imager 
would be extremely useful to extend current reddening maps and 
to increase their angular resolution. 
This is mandatory to improve the intrinsic accuracy of the 
CMD, and in turn the empirical evaluations of cluster 
parameters. Finally, we also mention that new CCD data 
are also necessary to estimate the secondary period of 
RR Lyrae affected by amplitude modulation, and therefore 
to constrain on a quantitative basis the occurrence and 
the intimate nature of the {\em Blazhko effect}.

It is a pleasure to thank S. Covino for sending us BV data of 
cluster stars and G. Clementini for BVI data of field RR Lyrae 
variables. We also thank S. Cassisi and M. Marconi for providing us 
evolutionary and pulsational predictions in advance of publication.  
We wish also to acknowledge an anonymous referee for his/her  
detailed suggestions and helpful comments that improved both the 
content and the readability of this paper.
We are grateful to G. Kovacs for sending us his code for the Fourier 
analysis, F. Caputo for a critical reading of an early draft of this 
manuscript, C. Cacciari and L. Pulone for useful discussions on 
photometric calibrations, and M. Dolci for enlightening suggestions 
concerning IDL routines. This work was supported by MURST/COFIN~2000 
under the project: "Stellar Observables of Cosmological Relevance"

\appendix 
\section{Comments on individual variables}

V6- Our data do not cover the phases around minimum light. Therefore 
current amplitude estimates, that are larger than estimated by C84, 
are uncertain. However, the light curve published by L02 supports 
the hypothesis that this object presents an amplitude modulation 
({\em Blazhko effect}), since our V magnitude close to the maximum light 
is roughly 0.2 mag brighter than measured by L02. The difference 
seems real, since current calibration does agree with the calibration 
provided by L02.   

V19- The difference in the luminosity amplitudes between current and 
C84 estimates (about 0.3 mag) suggests that this variable is a 
Blazhko candidate. However, the light curve given by C84 does not 
show a well-defined maximum, and therefore the amplitude is slightly 
uncertain. 

V23- The light curve of this object is presented here for the first 
time. Both the period (recomputed) and the amplitudes agree quite 
well with the estimates provided by L02.

V26- It is a good Blazhko candidate, since the difference in the B 
amplitude between current and C84 estimate is of the order of 0.35 mag. 
The position of this object in the Bailey plane is also peculiar and 
supports previous hypothesis. 

V31- The difference between current and C84 B amplitude is quite large 
and roughly equal to 0.6 mag. However, the amplitude provided by C84 is 
affected by a large uncertainty, and therefore the change could not be 
real. The comparison between current mean V magnitude and the mean 
magnitude provided by C84 shows that the latter one is 0.12 mag brighter. 
This evidence suggest that it might be a blend in the C84 photometry.    

V36- The period of this object was estimated and the new value is 
0.482143 days, while the old one was 0.4757433 (C84). However, 
current and C84 amplitudes do agree within the errors.
Moreover, the comparison between current mean V magnitude and the 
mean magnitude provided by C84 shows that the latter one is 0.2 mag
brighter. This evidence suggests that it might be a blend in the C84 
photometry (see her Table IV). 

V37- This variable is the reddest object ($(B-V)_0\approx0.40$ mag) 
in our sample and presents a very low luminosity amplitude ($A_B=0.35$). 
According to its position in the CM diagram it could be located close 
to the red edge, but its period is 0.577 days, while several $RR_{ab}$ 
stars present periods longer than 0.62 days. 
However, the reddening correction for this variable increases from 
E(B-V)=0.24 to E(B-V)=0.27 if we adopt the B amplitude estimated by
C84 ($A_B=0.97$ mag). This change might account for previous 
peculiarities, since V37 becomes slightly bluer and roughly 0.1 mag 
brighter. This object is a candidate Blazhko RR Lyrae. 
   
V41- The light curve of this object is presented here for the first time. 
The new period 0.6650 days is quite similar to the estimate provided by 
SH73 (0.66 days).

V45- This variable is the object for which the dereddened colors derived 
through the KW01 relations present the largest discrepancy with current 
estimates. The difference could be due to our Fourier amplitudes. In fact, 
the Fourier series do not properly fit the light curve around the minimum. 
Unfortunately, this variable is not included in the C84 sample, and 
therefore we cannot assess whether it presents any peculiarity in the 
luminosity amplitude. 

V47- We estimated the period of this variable and we found that the
new value P=0.5164 days is quite similar to the period derived by C84, i.e.
P=0.5212 days. We also found an alternative period of 0.3414 days, but 
the shape of the light curve (see Fig. 17) is typical of a $RR_{ab}$ star. 
However, current amplitude ($A_B=0.50$) is too small for a $RR_{ab}$ with 
this period. This peculiarity suggests that this object might be a Blazhko 
RR Lyrae and supports the former classification by SH73 and C84.

V48- It is one of the two first overtone variables in common with C84 
and the difference in the B amplitude is larger than 0.3 mag.
However, current and C84 light curves are affected by a large scatter. 
This empirical evidence, once confirmed, would imply that V48 is the 
first cluster $RR_c$ that shows the {\em Blazhko effect}.   

V51- The difference between current and C84 B amplitude is quite large,  
but its light curve presents a large scatter, probably because it is 
located close to the edge of the frame. The evidence that this variable 
is affected by {\em Blazhko effect} is not firm.

V58- The difference between current and C84 B amplitude is roughly 
0.4 mag, and the phase coverage along the light curves is good.  
This variable is a good candidate to be {\em Blazhko RR Lyrae}. 
Note that this object is somehow peculiar, and indeed current mean
V magnitude is 0.15 mag brighter than the mean magnitude provided
by C84. 

V50, V76, V80, V90, V92, V1405, V2710- The light curves of these 
objects are presented here for the first time. Their periods are: 
0.5338, 0.52577, 0.588, 0.6024, 0.54047, 0.3346, 0.548 days.
  
V79- The B magnitude roughly changes from 13.7 to 13.6, while the 
V magnitude changes from 12.19 to 12.11 (see Fig. 18). We note that 
the variability of this object was also suggested by GW98.  The location 
of this variable in the CMD suggests that it could be a semiregular 
variable. Unfortunately, the time interval covered by our observations 
does not allow us to estimate the pulsation period.

%

\newpage
 
\figcaption{Instrumental magnitudes as a function of the instrumental 
v-i (top) and b-v (middle, bottom) color. Instrumental magnitudes were 
transformed into standard BVI magnitudes according to the photometric 
data collected by Stetson (2000) and available at the following web 
page: cadcwww.hia.nrc.ca/standards/}

\figcaption{Top: comparison between the (V,V-I) CMD provided by 
Rosenberg et al. (2000) and the current cluster mean line. 
Bottom: difference between present V magnitudes (open circles) 
and (B-V) colors (filled circles) for a sample of bright stars in 
common with CO97.}

\figcaption{B (stars), V (filled circles), and I (open circles) 
light curves versus the pulsational phase for RR Lyrae in NGC~3201. 
The pulsational phase was shifted in such a way that the phase of 
maximum light is equal to 0.5. Variable identifications are labeled.}

\figcaption{B and V light curves of $RR_{ab}$ variables with periods 
close to 0.5 days and poor phase coverage close to the phase of 
minimum and/or  maximum light. The solid lines display the Layden's 
template adopted to fit empirical data. See text for further details.}

\figcaption{Difference in the mean $<V>$ (top) and $<B>$ (bottom) 
magnitude for variables in common with C84. Triangles show first 
overtones, while circles $RR_{ab}$ stars. Diamonds display 
fundamental variables with poor phase coverage.}

\figcaption{Difference in the V (top) and in the B (bottom) amplitude 
versus period for variables in common with C84.  Variables that show 
amplitude variations larger than 0.10 mag in the V band and of 0.15 mag 
in the B band (dashed lines) have been labeled. The symbols are the 
same as in Fig. 5.}   

\figcaption{Difference in the mean $<B>$ magnitude 
($<B> = [B_{max}+B_{min}]\times0.5$) for variables in common with SH73.}   

\figcaption{Top: difference in the B amplitude versus the pulsation period 
for variables in common with SH73. Variables that show amplitude variations 
larger than 0.45 mag have been labeled. The symbols are the same as in 
Fig. 5. Middle: same as the top panel, but the difference is referred to 
the minimum in luminosity. Bottom: same as the middle panel, but for the 
maximum in luminosity.}  

\figcaption{Bailey diagram for RR Lyrae in our sample. Symbols are the 
same as in Fig. 5. Solid lines connect current amplitudes with the 
amplitudes estimated by C84 (open circles) for variables that show 
variations larger than 0.10 mag in the V (top) and of 0.15 mag in the 
B (bottom) band.}   

\figcaption{Difference for RR Lyrae in our sample between E(B-V) color 
excesses estimated using the PACZ relation and the W90 relation (top), 
the KW01 relation (middle), as well as the ACZ relation (bottom). 
Variables that show amplitude modulation
in the B and in the V band are marked with a cross. The open circle 
in the middle panel marks the position of the variable V45. Variables 
characterized by a poor phase coverage were not included.} 

\figcaption{Same as Fig. 10, but the difference refers to the E(V-I) 
color excesses estimated using the PAC relation and the M95 relation 
(top) as well as the KW01 relation (bottom). Symbols are the same as 
in Fig. 10.}  

\figcaption {Comparison in the $V_0$,$(B-V)_0$ (left) and in the 
$V_0$,$(V-I)_0$ (right) CMD between theory and observations. 
The solid line shows the ZAHB,  while the dashed line the exhaustion 
of central He burning for [M/H]=-1.13. Theoretical predictions were 
plotted by adopting a distance modulus of $13.32\pm0.06$. 
HB stars and $RR_c$ variables 
were dereddened by smoothing with a spline the reddening maps based 
on $RR_{ab}$ stars. The error bars display the uncertainties on 
magnitude and colors due to errors on individual reddening estimates.  
Symbols are the same as in Fig. 5, open squares display HB stars.}

\figcaption{Same as Fig. 12, but for RR Lyrae variables. Solid and 
dashed lines show predicted fundamental and first overtone instability 
edges respectively. Adopted stellar mass and chemical composition are 
labeled. The comparison was performed by adopting a distance modulus 
of $13.32\pm0.06$. Symbols are the same as in Fig. 5.} 

\figcaption{Top: comparison in the Bailey diagram between RR Lyrae in 
NGC~3201, in M~5 (open circles), and in IC~4499 (open squares) respectively. 
Solid and dashed lines display predicted amplitudes 
for pulsation models constructed by adopting different luminosity levels
(see labeled values). Middle: same as top, but for V amplitudes. 
Note that top and the middle panels show 5 $RR_c$ stars, since we 
also included the two extra $RR_c$ variables observed by C84. 
Bottom: same as top, but for I amplitudes.}

\figcaption {Reddened (top) and dereddened (bottom) (V,B-V) and (V,V-I) 
CMDs. Individual reddening evaluations were obtained by adopting the 
reddening maps based on $RR_{ab}$ colors. As a consequence, the bottom 
panels only display the subsample of cluster stars located inside 
these maps. Note the substantial decrease in the thickness of RGB stars, 
as well as in the subgiant and in the turn-off region. The solid lines
in the bottom panels show the fiducial lines for M~3 according to 
Ferraro et al. (1997, V,B-V) and Johnson \& Bolte (1998, V,V-I). 
To overplot the two sets of fiducial lines we applied a magnitude 
shift of -1.85 and -1.8 respectively.}

\figcaption {Comparison between predicted and empirical 
$\Delta{V}_{HB}^{Bump}$ values as a function of global metallicity. 
Long-dashed, solid, and dashed lines show theoretical predictions 
for three different cluster ages (see labeled values). 
To avoid systematic uncertainties that could affect current metallicity 
scales (Rutledge et al. 1997) we only plotted GGCs for which are available 
spectroscopic measurements of iron and $\alpha$-element abundances.}   

\figcaption{B (left) and V (right) band light curves of V47 phased by 
adopting two different periods namely 0.5164 and 0.3414 days.} 

\figcaption{V (top) and B (bottom) magnitudes as a function of the Julian 
day for the suspect red variable V79. This object is located close to the 
tip of the RGB.}

\end{document}


\tablewidth{0pt} 
\begin{deluxetable}{lllrccccccccc}
\tablecaption{Pulsational properties of RR Lyrae in NGC~3201}\label{tbl-1}
\tablehead{
\colhead{ID \tablenotemark{a}}&
\colhead{Type}& 
\colhead{Period\tablenotemark{b}}&
\colhead{Ref.\tablenotemark{c}}&
\colhead{$B_m$\tablenotemark{d}}&
\colhead{$V_m$\tablenotemark{d}}&
\colhead{$I_m$\tablenotemark{d}}& 
\colhead{$<B>$\tablenotemark{e}}&
\colhead{$<V>$\tablenotemark{e}}&
\colhead{$<I>$\tablenotemark{e}}&
\colhead{A(B)\tablenotemark{f}}& 
\colhead{A(V)\tablenotemark{f}}& 
\colhead{A(I)\tablenotemark{f}}
}
\startdata
1         & ab & 0.6048761 & SH73 &15.51 & 14.83 & 13.95 & 15.45 & 14.79 &  13.94 &  1.15  & 0.91 & 0.57 \nl  
2$^g$     & ab & 0.5326722 & C84&15.47 & 14.87 & \ldots& 15.35 & 14.82 &  \ldots&  1.25  & 1.00 & \ldots\nl  
3$^h$     & ab & 0.5993792 & C84&15.60 & 14.90 & 13.97 & 15.54 & 14.87 &  13.96 &  1.11  & 0.85 & 0.56\nl  
4         & ab & 0.6300096 & C84&15.50 & 14.81 & 13.87 & 15.43 & 14.77 &  13.86 &  1.29  & 1.02 & 0.69\nl 
5         & ab & 0.501536  & SH73 &15.38 & 14.78 & 13.99 & 15.27 & 14.71 &  13.97 &  1.52  & 1.20 & 0.78\nl   
6$^{g,h}$ & ab & 0.5250936 & C84&15.37 & 14.76 & \ldots& 15.21 & 14.67 &  \ldots&  1.8   & 1.40 & \ldots\nl 
7         & ab & 0.6303322 & SH73 &15.33 & 14.68 & \ldots& 15.31 & 14.67 &  \ldots&  0.63  & 0.50 & \ldots\nl
8         & ab & 0.6286573 & C84&15.42 & 14.72 & 13.90 & 15.41 & 14.71 &   13.89&  0.60  & 0.48 & 0.35\nl  
9         & ab & 0.5267087 & C84&15.45 & 14.81 & 14.03 & 15.36 & 14.76 &   14.02&  1.45  & 1.10 & 0.71\nl 
11        &  c & 0.2990490 & C84&15.29 & 14.78 & 14.15 & 15.26 & 14.76 &   14.15&  0.70  & 0.54 & 0.34\nl
12$^g$    & ab & 0.4955547 & C84& 15.44& 14.78 & \ldots& 15.33 & 14.71 &  \ldots&  1.45  & 1.2  & \ldots\nl
13        & ab & 0.5752145 & SH73 &15.54 & 14.85 & 14.03 & 15.48 & 14.81 &   14.02&  1.15  & 0.96 & 0.56\nl
17        & ab & 0.5655773 & SH73 &15.39 & 14.77 & 13.93 & 15.34 & 14.74 &   13.92&  1.11  & 0.88 & 0.54\nl   
18$^g$    & ab &  0.5395   & L02& 15.35& 14.74 & \ldots& 15.29 & 14.70 &  \ldots&  1.13  & 0.9  & \ldots\nl
19$^h$    & ab & 0.5250201 & SH73 &15.42 & 14.76 & \ldots& 15.36 & 14.74 &  \ldots&  1.13  & 0.90 & \ldots\nl
20$^g$    & ab &  0.5291045& C84& 15.38& 14.75 & \ldots& 15.30 & 14.71 &  \ldots&  1.3   & 0.98 & \ldots\nl
21        & ab & 0.5666280 & C84&15.50 & 14.84 & 14.00 & 15.46 & 14.82 &   13.99&  1.05  & 0.73 & 0.48\nl
22        & ab & 0.6059882 & C84&15.39 & 14.73 & 13.91 & 15.34 & 14.70 &   13.90&  1.03  & 0.81 & 0.67\nl
23        & ab & 0.586     & PW &15.44 & 14.78 & 13.91 & 15.40 & 14.76 &   13.90&  0.97  & 0.73 & 0.51\nl
25        & ab & 0.5148010 & C84&15.38 & 14.78 & 13.98 & 15.27 & 14.73 &   13.96&  1.43  & 1.03 & 0.70\nl
26$^h$    & ab & 0.5689721 & C84&15.63 & 14.93 & 14.07 & 15.55 & 14.89 &   14.05&  1.53  & 1.18 & 0.62\nl
28        & ab & 0.5786766 & SH73 &15.52 & 14.84 & 13.98 & 15.43 & 14.81 &   13.96&  1.37  & 1.04 & 0.70\nl
31$^h$    & ab & 0.5197267 & C84&15.60 & 14.95 & 14.04 & 15.56 & 14.93 &  14.03 &  0.98  & 0.73 & 0.47\nl
32        & ab & 0.5611656 & SH73 &15.41 & 14.80 & 13.92 & 15.29 & 14.72 &  13.90 &  1.62  & 1.24 & 0.80\nl
34$^h$    & ab & 0.4678900 & C84&15.37 & 14.80 & 14.03 & 15.21 & 14.75 &  14.00 &  1.71  & 1.27 & 0.84\nl
35        & ab & 0.6155244 & C84&15.42 & 14.74 & 13.87 & 15.38 & 14.73 &  13.86 &  0.79  & 0.62 & 0.40\nl
36        & ab & 0.482143  & PW &15.29 & 14.74 & 13.98 & 15.23 & 14.71 &  13.97 &  1.15  & 0.85 & 0.53\nl
37$^h$    & ab & 0.5772897 & C84&15.37 & 14.74 & 13.92 & 15.37 & 14.74 &  13.92 &  0.35  & 0.21 & 0.16\nl
38$^{g,h}$&ab &  0.5091250& C84&15.44 & 14.82 & \ldots& 15.35 & 14.78 &  \ldots&  1.38  & 1.03 & \ldots\nl
39        & ab & 0.4832092 & SH73 &15.47 & 14.85 & 14.04 & 15.36 & 14.79 &  14.01 &  1.56  & 1.22 & 0.77\nl
40        & ab & 0.6421096 & C84&15.50 & 14.79 & 13.91 & 15.48 & 14.78 &  13.91 &  0.59  & 0.43 & 0.29\nl
41        & ab & 0.665     &  PW&15.52 & 14.82 & 13.85 & 15.50 & 14.81 &  13.85 &  0.56  & 0.45 & 0.28\nl
44        & ab &  0.6107344& SH73 &15.44 & 14.76 & 13.86 & 15.41 & 14.75 &  13.86 &  0.89  & 0.66 & 0.41\nl
45        & ab &  0.5374165& SH73 &15.61 & 14.94 & 14.03 & 15.48 & 14.87 &  14.01 &  1.73  & 1.26 & 0.83\nl
47$^h$    & ab & 0.5164   & PW & 15.23 & 14.67 & 13.95 & 15.22 & 14.66 &  13.95 &  0.50  & 0.33 & 0.20\nl
48$^h$    &  c & 0.3412136 & C84&15.27 & 14.68 & \ldots& 15.24 & 14.67 &  \ldots&  0.87  & 0.54 & \ldots\nl
49        & ab &  0.5815089& C84&15.37 & 14.73 & 13.93 & 15.29 & 14.69 &  13.91 &  1.33  & 0.99 & 0.64\nl 
50$^g$    & ab &  0.5338   &  PW&15.26 & 14.66 & \ldots& 15.16 & 14.60 &  \ldots&  1.45  & 1.15 & \ldots\nl
51$^h$    & ab &  0.5206220& C84&15.44 & 14.76 & \ldots& 15.34 & 14.70 &  \ldots&  1.50  & 1.22 & \ldots\nl  
56        & ab &  0.5903376& C84&15.62 & 14.92 & 13.97 & 15.58 & 14.89 &  13.96 &  1.08  & 0.79 & 0.51\nl
57        & ab &  0.5934337& C84&15.52 & 14.85 & \ldots& 15.47 & 14.83 &  \ldots&  0.95  & 0.71 & \ldots\nl  
58$^h$    & ab & 0.6220418 & C84&15.42 & 14.73 & 13.85 & 15.39 & 14.70 &  13.84 &  0.95  & 0.72 & 0.43\nl
71        & ab & 0.6011859 & SH73 &15.39 & 14.71 & \ldots& 15.35 & 14.69 &  \ldots&  0.99  & 0.68 & \ldots\nl 
73        & ab & 0.5199500 & SH73 &15.45 & 14.80 & 13.98 & 15.32 & 14.73 &  13.96 &  1.64  & 1.25 & 0.83\nl  
76$^g$    & ab &  0.52577  &  PW&15.42 & 14.80 & \ldots& 15.32 & 14.75 &  \ldots&  1.42  & 1.13 & \ldots\nl
77        & ab & 0.5676648 & SH73 &15.34 & 14.71 & 13.91 & 15.27 & 14.67 &  13.89 &  1.16  & 0.89 & 0.64\nl
78        & ab & 0.514     & SH73 &15.45 & 14.82 & 14.02 & 15.39 & 14.79 &  14.01 &  1.00  & 0.81 & 0.50\nl
80        & ab & 0.588     & PW &15.40 & 14.76 & 13.87 & 15.34 & 14.73 &  13.85 &  1.16  & 0.85 & 0.56\nl
83        & ab & 0.5451998 & C84&15.46 & 14.84 & 13.95 & 15.36 & 14.78 &   13.93&  1.50  & 1.17 & 0.79\nl 
90        & ab & 0.6024    & PW &15.33 & 14.69 & 13.87 & 15.26 & 14.65 &  13.85 &  1.24  & 0.95 & 0.61\nl
92$^g$    & ab &  0.54047  & PW &15.38 & 14.75 & \ldots& 15.27 & 14.70 &  \ldots&  1.47  & 1.1  & \ldots\nl   
1405      &  c & 0.3346    & PW &15.33 & 14.76 & 14.04 & 15.31 & 14.75 &   14.04&  0.51  & 0.40 & 0.27\nl   
2710      & ab & 0.548     & PW &15.39 & 14.77 & 13.95 & 15.30 & 14.72 &  13.93 &  1.38  & 1.06 & 0.72\nl
\enddata
\tablenotetext{a}{Variable identification. The number is from Sawyer-Hogg (1973) except for the new variables 
for which the Lee's (1971) designation has been adopted. 
\hspace*{0.5mm}$^b$ Pulsational period (days). 
\hspace*{0.5mm} $^c$ References for period estimates (C84: Cacciari 1984; SH73: Sawyer Hogg 1973; 
PW: present work; L02: Layden et al. 2002).
\hspace*{0.5mm} $^d$ Mean magnitude-weighted B, V, and I magnitudes.
\hspace*{0.5mm} $^e$ Mean intensity-weighted B, V, and I magnitudes. 
\hspace*{0.5mm} $^f$ Luminosity amplitude in the B, V, and I band (mag). 
\hspace*{0.5mm} $^g$ Variables whose B and V light curves were fitted with 
the Layden's template (Layden 1998).  
\hspace*{0.5mm} $^h$ Variables that show B and V amplitude modulation. 
}
\end{deluxetable}

\clearpage
\begin{deluxetable}{lccccc}
\tablecaption{Mean $(B-V)$ colors and reddenings for RR Lyrae in NGC3201}\label{tbl-2}
\tablehead{
\colhead{ID}&
\colhead{$(B-V)_m$\tablenotemark{a}}& 
\colhead{$(B-V)_{min}$\tablenotemark{b}}&
\colhead{$E(B-V)_{W90}$\tablenotemark{c}}&
\colhead{$E(B-V)_{PACZ}$\tablenotemark{d}}&
\colhead{$(B-V)_{0,PACZ}$\tablenotemark{e}}
}
\startdata 
   1     & 0.69 & 0.75 & 0.35 & 0.33 & 0.36\nl 
 3$^f$ & 0.69 & 0.75 & 0.36  &0.34 & 0.35\nl
   4     & 0.69 & 0.77 & 0.37  &0.34 & 0.35\nl
   5     & 0.61 & 0.70 & 0.33 & 0.29 & 0.32\nl
   7     & 0.65   &  0.71 & 0.31 & 0.27 & 0.38\nl
   8     &  0.70   &  0.74   &  0.34 & 0.32 & 0.38\nl
   9     &  0.64   &  0.72   &  0.35 & 0.31 & 0.33\nl
  13     &  0.69   &  0.78   &  0.40 & 0.34 &  0.35\nl
  17     &  0.62   &  0.70   &  0.32 & 0.27   &  0.35\nl
19$^f$   &  0.65 & 0.73 & 0.36 & 0.31   &  0.34\nl
  21     &  0.66 & 0.68 & 0.30 & 0.31   &  0.35\nl
  22     &  0.66 & 0.74 & 0.35 & 0.30   &  0.36\nl
  23     &  0.66 & 0.73 & 0.34 & 0.30 & 0.36\nl
  25     &  0.59 & 0.70 & 0.33 & 0.27 & 0.32\nl
  26$^f$ &  0.70 & 0.77 & 0.39 & 0.37 & 0.33\nl
  28     & 0.67 & 0.77   &  0.39 & 0.33 & 0.34\nl
  31$^f$ & 0.65 & 0.76   &  0.39 & 0.30 & 0.35\nl
  32     & 0.61 & 0.70 & 0.32 & 0.29 & 0.32\nl
  34$^f$ & 0.57 & 0.67 & 0.31 & 0.27 & 0.30\nl
  35     & 0.67 & 0.72 & 0.33  &0.30 & 0.37\nl
  36     & 0.55 & 0.65 & 0.29  &0.22   &  0.33\nl
  37$^f$ & 0.63 & 0.67 & 0.29 & 0.24   &  0.39\nl
  39     & 0.62 & 0.72 & 0.36 & 0.31 & 0.31\nl
  40     & 0.71 & 0.76 & 0.36 & 0.32 & 0.39\nl
  41     & 0.70 & 0.74 & 0.33 & 0.31 & 0.39\nl
  44     & 0.68 & 0.74 & 0.35 & 0.31 & 0.37\nl
  45     & 0.67 & 0.74 & 0.37 & 0.35 & 0.31\nl
  47$^f$ & 0.56 & 0.61 & 0.24 & 0.19 & 0.37\nl
  49     & 0.64 & 0.74 & 0.35 & 0.30 & 0.34\nl
  51$^f$ & 0.68 & 0.75 & 0.38 & 0.36 & 0.32\nl
  56     & 0.70 & 0.75 & 0.36 & 0.35 & 0.35\nl
  57     & 0.66 & 0.74 & 0.35 & 0.30 & 0.36\nl
  58$^f$ & 0.70 & 0.75 & 0.35 & 0.33 & 0.37\nl
  71     &  0.68 & 0.75 & 0.36 & 0.32 & 0.36\nl
  73     & 0.65 & 0.76 & 0.39 & 0.33 & 0.32\nl
  77     &  0.63 & 0.70 & 0.32 & 0.28 & 0.35\nl
  78     & 0.63 & 0.66 & 0.29 & 0.28 & 0.35\nl
  80     & 0.64 & 0.72 & 0.33 & 0.29 & 0.35\nl
  83     & 0.62 & 0.71 & 0.33 & 0.29 & 0.33\nl
  90     & 0.64 & 0.71 & 0.32 & 0.29 & 0.35\nl
2710     & 0.62 & 0.71 & 0.33 & 0.29 & 0.33\nl
\enddata
\tablenotetext{a}{Mean magnitude-weighted (B-V) color. 
\hspace*{0.5mm} $^b$ Mean color at minimum phase, i.e. $(0.5 \le \Phi \le 0.8)$.
\hspace*{0.5mm} $^c$ Reddening estimate based on the Sturch (1966) method 
according to the Walker's calibration (Walker 1990).
\hspace*{0.5mm} $^d$ Reddening estimate according to the PACZ relation 
(see text). 
\hspace*{0.5mm} $^e$ Mean (B-V) color dereddened using the PACZ relation.
\hspace*{0.5mm} $^f$ Variables that show B and V amplitude modulation. 
} 
\end{deluxetable}

\clearpage
\begin{deluxetable}{lccccccc}
\tablecaption{Reddening determinations of selected GGCs with RR Lyrae}\label{tbl-1}
\tablehead{
\colhead{Cluster\tablenotemark{a}}&
\colhead{[Fe/H]\tablenotemark{b}}&
\colhead{E(B-V)\tablenotemark{c}}&
\colhead{$\Delta$E(B-V)\tablenotemark{d}}&
\colhead{E(B-V)\tablenotemark{e}}&
\colhead{E(V-I)\tablenotemark{f}}&
\colhead{$RR_{ab}$\tablenotemark{g}}&
\colhead{Ref.\tablenotemark{h}}
}
\startdata
NGC~1851     & -1.36 & 0.02 & 0.02 & $0.04\pm0.01$ & $0.02\pm0.01$  &7 & 1 \\   
NGC~4590/M~68& -2.09 & 0.05 & 0.01 & $0.05\pm0.01$ & $0.06\pm0.02$  &7 & 2 \\  
NGC~5466     & -2.22 & 0.00 & 0.02 & $0.01\pm0.01$ & \ldots         &7 & 3 \\   
NGC~5904/M~5 & -1.40 & 0.03 & 0.01 & $0.04\pm0.03$ & $0.05\pm0.02$  &21,33\tablenotemark{i} &4,5,6,7\\ 
NGC~6341/M~92& -2.24 & 0.02 & 0.02 & $0.04\pm0.02$ & \ldots         &6 &8 \\ 
NGC~6441     & -0.5$\star$& 0.44 & 0.17 & $0.53\pm0.14$ & \ldots   & 15 & 9 \\ 
NGC~7006     & -1.59 & 0.05 & 0.03 & $0.01\pm0.03$ & \ldots         &15 &10 \\  
NGC~7089/M~2 & -1.62 & 0.06 &-0.02 & $0.01\pm0.01$ & \ldots         &12 &11 \\ 
Ruprecht~106 & -1.7$\star$&0.20 & -0.03 & $0.15\pm0.02$ & \ldots   & 13 & 12 \\ 
\enddata
\tablenotetext{a}{Cluster name. 
\hspace*{0.5mm} $^b$ Cluster metallicity according to ZW84. Clusters   
marked with a $\star$ are from Harris (1996).   
\hspace*{0.5mm} $^c$ Cluster reddening based on light emitted by cluster 
members (Dutra \& Bica 2000).  
\hspace*{0.5mm} $^d$ Difference between the cluster reddening based on 
far-infrared dust emission (Schlegel et al. 1998) and the reddening 
listed in the previous column (Dutra \& Bica 2000).  
\hspace*{0.5mm} $^e$ Cluster reddening and standard deviation based on 
current B,V photometry and the PACZ relation.   
\hspace*{0.5mm} $^f$ Cluster reddening and standard deviation based on 
current V,I photometry and the PAC relation.   
\hspace*{0.5mm} $^g$ Number of fundamental RR Lyrae adopted to estimate the 
mean cluster reddening.  
\hspace*{0.5mm} $^h$ Source of the photometry for cluster RR Lyrae:
1) Walker 1998;  2) Walker 1994; 3) Corwin et al. 1999; 
4) Storm et al. 1991; 5) Reid 1996; 
6) Brocato et al. 1996; 7) Caputo et al. 1999; 8) Carney et al. 1992; 
9) Pritzl et al. 2001;
10) Wehlau et al. 1999; 
11) Lee \& Carney 1999;
12) Kaluzny et al. 1995.  
\hspace*{0.5mm} $^i$ The former value refers to B-V colors, while the 
latter to V-I colors.  
}
\end{deluxetable}

\begin{deluxetable}{rllccccccc}
\tablecaption{The main pulsational properties of field $RR_{ab}$ stars
with BVI data}\label{tbl-4}
\tablehead{
\colhead{ID}& 
\colhead{$[Fe/H]$}&
\colhead{$\log P$}&
\colhead{$A(B)$}& 
\colhead{$<B-V>$}&
\colhead{$<V-I>$}&
\colhead{$A(V)$}&
\colhead{$E(B-V)$}&
\colhead{$E(V-I)$}&  
\colhead{Ref.\tablenotemark{a}}
}
\startdata
SW And   &  -0.15 &  -0.35432 & 1.27 &  0.430 &   0.537  &   0.95  &  0.06  & 0.07  & 1\nl     
X Ari    &  -2.40 &  -0.1863  & 1.26 &  0.48  &    0.71  &    1.01 &  0.14  & 0.17  & 2\nl
RR Cet   &  -1.25 &  -0.25725 & 1.21 &  0.366 &   0.525  &   0.93  &  0.03  & 0.04  & 1\nl    
UU Cet   &  -1.28 &  -0.2175  & 0.85 &  0.384 &   0.555  &   0.65  &  0.015 & 0.02  & 3\nl
SU Dra   &  -1.6  &  -0.18018 & 1.26 &  0.344 &   0.512  &   0.97  &  0.01  & 0.01  & 1\nl    
SW Dra   &  -1.12 &  -0.2444  & 1.15 &  0.36  &    0.515 &    0.90 &  0.015 & 0.02  & 4\nl
RX Eri   &  -1.4  &  -0.23118 & 1.14 &  0.413 &   0.579  &   0.88  &  0.05  & 0.06  & 1\nl  
SS For   &  -0.94 &  -0.3050  & 1.65 &  0.30  &    0.45  &    1.3  &  0.015 & 0.02  & 4\nl
RR Gem   &  -0.20 &  -0.40087 & 1.62 &  0.393 &   0.501  &   1.21  &  0.075 & 0.10  & 1\nl      
V Ind    &  -1.50 &  -0.3191  & 1.35 &  0.338 &    0.488 &  1.05   &  0.015 & 0.02  & 3\nl  
RR Leo   &  -1.15 &  -0.34449 & 1.64 &  0.329 &   0.472  &   1.30  &  0.05  & 0.02  & 1\nl       
TT Lyn   &  -1.35 &  -0.22371 & 0.92 &  0.378 &   0.548  &   0.70  &  0.01  & 0.01  & 1\nl     
AV Peg   &   0.00 &  -0.40852 & 1.41 &  0.419 &   0.525  &   1.03  &  0.07  & 0.09  & 1\nl      
RV Phe   &  -1.69 &  -0.2245  & 0.90 & 	0.37  &   0.53   &    0.65 &  0.015 & 0.02  & 4\nl 
TU Uma   &  -1.25 &  -0.25365 & 1.22 &  0.36  &   0.517  &   0.94  &  0.02  & 0.025 & 1\nl      
V440 Sgr &  -1.40 &  -0.3235  & 1.6  &  0.40  &   0.55   &   1.2   &  0.12  & 0.15  & 4\nl
W Tuc    &  -1.57 &  -0.1923  & 1.4  &  0.323 &   0.482  &    1.10 &  0.015 & 0.02  & 3\nl
UU Vir   &  -0.55 &  -0.32275 & 1.5  &  0.335 &   0.440  &   1.17  &  0.01  & 0.01  & 1\nl
\enddata 
\tablenotetext{a}{References: 1) Liu \& Janes 1990; 2) Fernley et al. 1989;
3) Clementini et al. 1990; 4) Cacciari et al. 1987 }  
\end{deluxetable}

\clearpage
\begin{deluxetable}{lccccc}
\tablecaption{Mean $(V-I)$ colors and reddenings for RR Lyrae in NGC3201}\label{tbl-3}
\tablehead{
\colhead{ID}&
\colhead{$(V-I)_m$\tablenotemark{a}}&
\colhead{$(V-I)_{min}$\tablenotemark{b}}&
\colhead{$E(V-I)_{PAC}$\tablenotemark{c}}&
\colhead{$E(V-I)_{M95}$\tablenotemark{d}}&
\colhead{$(V-I)_0$\tablenotemark{e}}  
}
\startdata
     1  & 0.87    & 0.96 & 0.37   & 0.38   &  0.49\nl
 3$^g$& 0.93    & 1.01 & 0.42   & 0.43   &  0.50\nl
     4  & 0.94    & 1.04 & 0.43   & 0.46   &  0.48\nl
     5  & 0.78    & 0.90 & 0.32   & 0.32   &  0.44\nl          
     8  & 0.83$^f$& 0.86 &0.28$^f$& 0.28   &  0.53\nl
     9  & 0.78$^f$& 0.88 & 0.30   &0.30$^f$&  0.46\nl
    13  & 0.82$^f$& 0.90 & 0.32   &0.32$^f$&  0.49\nl
    17  & 0.84    & 0.90 & 0.34   & 0.32   &  0.50\nl          
    21  & 0.84    & 0.93 & 0.33   & 0.35   &  0.51\nl
    22  & 0.82$^f$& 0.89 & 0.30   &0.31$^f$&  0.51\nl
    23  & 0.87    & 0.91 & 0.35   & 0.33   &  0.52\nl
    25  & 0.81    & 0.86 & 0.33   & 0.28   &  0.48\nl
26$^g$  & 0.86    & 0.92 & 0.39   & 0.34   &  0.48\nl
    28  & 0.87    & 0.97 & 0.38   & 0.39   &  0.47\nl
31$^g$  & 0.91    & 0.95 & 0.41   & 0.37   &  0.51\nl
    32  & 0.87    & 0.94 & 0.40   & 0.36   &  0.46\nl
34$^g$  & 0.77    & 0.88 & 0.33   & 0.30   &  0.43\nl
    35  & 0.88    & 0.88 & 0.35   & 0.30   &  0.55\nl
    36  & 0.76$^f$& 0.82 & 0.28   & 0.24   &  0.49\nl
37$^g$  & 0.82    & 0.85 & 0.27   & 0.27   &  0.57\nl
    39  & 0.81    & 0.90 & 0.35   & 0.32   &  0.44\nl
    40  & 0.88$^f$& 0.91 & 0.33   &0.33$^f$&  0.56\nl
    41  & 0.96    & 1.00 & 0.41   & 0.42   &  0.56\nl
    44  & 0.90    & 0.92 & 0.37   & 0.34   &  0.54\nl
    45  & 0.91    & 1.03 & 0.45   & 0.45   &  0.44\nl
47$^g$  & 0.72    & 0.79 & 0.20   & 0.21   &  0.51\nl
    49  & 0.80    & 0.86 & 0.31   & 0.28   &  0.49\nl         
    56  & 0.96    & 1.02 & 0.44   & 0.44   &  0.51\nl        
58$^g$  & 0.88    & 0.98 & 0.35   & 0.40   &  0.51\nl         
    73  & 0.82$^f$& 0.84 & 0.36   &0.26$^f$&  0.48\nl
    77  & 0.80    & 0.88 & 0.30   & 0.30   &  0.49\nl
    78  & 0.80$^f$& 0.86 & 0.31   & 0.28   &  0.50\nl
    80  & 0.89    & 0.95 & 0.38   & 0.37   &  0.51\nl
    83  & 0.89    & 0.96 & 0.42   & 0.38   &  0.47\nl
    90  & 0.82    & 0.90 & 0.32   & 0.32   &  0.49\nl
  2710  & 0.82    & 0.90 & 0.34   & 0.32   &  0.47\nl
\enddata
\tablenotetext{a}{Mean magnitude-weighted (V-I) color. 
\hspace*{0.5mm} $^b$ Mean color at minimum phase, i.e. $(0.5 \le \Phi \le 0.8)$.
\hspace*{0.5mm} $^c$ Reddening estimate based on the PAC relation. 
\hspace*{0.5mm} $^d$ Reddening estimate based on the empirical evidence that 
for RR Lyrae $(V-I)_{0,min}$ = $0.58\pm=0.03$ mag (Mateo et al. 1995).   
\hspace*{0.5mm} $^e$ Mean (V-I) color dereddened according to the reddenings 
listed in the fourth and in the fifth column (weighted average).
\hspace*{0.5mm} $^f$ Variables that present larger photometric uncertainty.  
\hspace*{0.5mm} $^g$ Variables that show B and V amplitude modulation. 
}
\end{deluxetable}

\clearpage
\begin{deluxetable}{lccccccccc}
\tablecaption{Fourier parameters\tablenotemark{a} and reddening estimates 
for RR Lyrae in NGC~3201}\label{tbl-6}
\tablehead{
\colhead{ID\tablenotemark{b}}&
\colhead{$A_1$\tablenotemark{c}}& 
\colhead{$R_{21}$\tablenotemark{d}}&
\colhead{$R_{31}$\tablenotemark{d}}&
\colhead{$\Phi_{21}$\tablenotemark{e}}&
\colhead{$\Phi_{31}$\tablenotemark{e}}&
\colhead{$E(B-V)$\tablenotemark{f}}&
\colhead{$E(V-I)$\tablenotemark{f}}&
\colhead{$(B-V)_0$\tablenotemark{g}}&
\colhead{$(V-I)_0$\tablenotemark{g}} 
}
\startdata
1     & 0.305 &  0.470 & 0.326   &  2.415  & 5.222  & 0.33 & 0.36 & 0.35   & 0.51 \nl
3$^h$ & 0.283 &  0.458 & 0.310   &  2.419  & 5.279  & 0.34 & 0.42 & 0.35   & 0.51 \nl
4     & 0.333 &  0.555 & 0.325   &  2.636  & 5.512  & 0.34 & 0.43 &  0.35  & 0.51 \nl
5     & 0.442 &  0.430 & 0.304   &  2.136  & 4.828  & 0.30 &  0.33&  0.30  & 0.45 \nl
7     & 0.193 & 0.360  & 0.109   &  2.132  & 6.025  & 0.28 &\ldots&   0.37 & 0.53 \nl
8     & 0.207 & 0.387  & 0.199   &  2.820  & 5.891  & 0.33 &  0.30& 0.37   & 0.53 \nl
9     & 0.357 & 0.553  & 0.316   &  2.320  & 4.850   & 0.31 & 0.30 &  0.33  & 0.48 \nl
11    & 0.269 &  0.196 & 0.067   &  3.024  & 5.982   &\ldots&\ldots&\ldots  &\ldots \nl
13    & 0.323 & 0.538  & 0.329   &  2.620  & 5.460   &  0.35& 0.32 &  0.34  & 0.50 \nl
17    & 0.311 & 0.503  & 0.310   &  2.340  & 5.058   & 0.27 & 0.34 &  0.34  & 0.50 \nl
19$^h$& 0.267 &  0.465 & 0.381   &  2.520  & 5.350   & 0.30 &\ldots&  0.35  & 0.51 \nl
21    & 0.255 &  0.487 & 0.311   &  2.465  & 5.287   &  0.30& 0.33 &  0.36  & 0.52 \nl
22    & 0.275 & 0.488  & 0.344   &  2.420  & 5.350   & 0.30 & 0.30 &  0.36  & 0.52 \nl
23    & 0.262 & 0.466  & 0.281   &  2.420  & 5.240   & 0.31 & 0.36 &  0.36  & 0.51  \nl
25    & 0.394 & 0.461  & 0.293   &  2.140  & 4.580   & 0.28 & 0.34 &  0.32  & 0.46 \nl
26$^h$& 0.337 &  0.619 & 0.375   &  2.656  & 5.678   &  0.35&  0.36&  0.34  & 0.50 \nl
28    & 0.360 & 0.522  & 0.321   &  2.346  & 4.997   &  0.33& 0.38 &  0.34  & 0.49 \nl
31$^h$& 0.282 & 0.475  & 0.291   &  2.455  & 5.359   & 0.31 & 0.41 &  0.34  & 0.50 \nl
32    & 0.441 & 0.478  & 0.318   &  2.320  & 4.980   & 0.30 & 0.41 &  0.32  & 0.46 \nl
34$^h$& 0.443 & 0.441  & 0.295   &  2.292  & 4.833   & 0.27 & 0.33 & 0.30   & 0.44 \nl
35    & 0.224 & 0.476  & 0.276   &  3.020  & 5.230   & 0.30 & 0.35 & 0.37   & 0.53 \nl
36    & 0.356 & 0.395  & 0.209   &  2.290  & 4.882   & 0.24 & 0.30 & 0.31   & 0.46 \nl
37$^h$& 0.072 & 0.447  & 0.286   &  2.380  & 4.990   & 0.23 & 0.26 &  0.40  & 0.57 \nl
39    & 0.494 & 0.410  & 0.254   &  2.120  & 4.860   & 0.34 & 0.38 &   0.28 & 0.42 \nl
40    & 0.171 & 0.418  & 0.210   &  2.554  & 5.656   & 0.33 & 0.34 &  0.38  & 0.55 \nl
41    & 0.174 & 0.415  & 0.209   &  2.670  & 5.520   & 0.32 & 0.41 &  0.38  & 0.55 \nl
44    & 0.248 & 0.426  & 0.291   &  2.430  & 5.360   & 0.32 & 0.38 &  0.36  & 0.52 \nl
45    & 0.372 & 0.529  & 0.559   &  1.690  & 4.435  & 0.32 & 0.40 & 0.35   &  0.51 \nl
47$^h$  & 0.143 & 0.116  & 0.067   &  0.886  & 3.102  &  0.19& 0.20 & 0.37 & 0.52 \nl
48$^h$& 0.236 & 0.138  & 0.094   &  3.514  & 6.985  &\ldots&\ldots&\ldots  &\ldots \nl
49    & 0.354 & 0.412  & 0.312   &  2.340  & 4.910   &  0.30& 0.32 &  0.34  &  0.49 \nl
51$^h$& 0.401 & 0.576  & 0.373   &  2.340  & 4.870   &  0.35&\ldots&  0.32  &  0.48 \nl
56    & 0.304 & 0.504  & 0.388   &  2.394  & 5.150   &  0.35& 0.44 &  0.36  &  0.52 \nl
57    & 0.249 & 0.483  & 0.395   &  2.420  & 5.190   & 0.30 &\ldots&  0.37  &  0.53 \nl
58$^h$& 0.252 & 0.506  & 0.312   &  2.366  & 5.247   & 0.33 & 0.35 &  0.37  &  0.53 \nl
71    & 0.253 & 0.434  & 0.291   &  2.725  & 5.831   & 0.32 &\ldots&  0.36  &  0.52 \nl
73    & 0.445 & 0.515  & 0.325   &  1.890  &  4.460   &  0.34& 0.37 & 0.31   & 0.46 \nl  
77    & 0.337 & 0.461  & 0.295   &  2.230  &  5.030   & 0.30 & 0.31 &  0.34  & 0.49 \nl
78    & 0.314  & 0.377 & 0.281   &  2.570  &  5.190   & 0.30 & 0.31 &  0.33  & 0.49 \nl
80    & 0.326 &  0.409 & 0.281   &  2.210  &  4.860   & 0.30 & 0.39 &  0.34  & 0.50 \nl
83    & 0.486 & 0.758  & 0.591   &  2.890  &  6.220   &  0.29& 0.39 & 0.34   & 0.50  \nl
90    & 0.323 & 0.501  & 0.336   &  2.430  &  5.248   & 0.29 & 0.31 & 0.35    & 0.51 \nl
1405  & 0.211 & 0.090  & 0.051   &  2.834  &  3.130   &\ldots&\ldots&\ldots  &\ldots \nl
2710  & 0.341 & 0.539  & 0.335   &  2.460  &  5.196   &  0.29& 0.33 & 0.34    &  0.49 \nl
\enddata
\tablenotetext{a}{ Fourier parameters estimated by fitting V-band light curves 
(see text).  
\hspace*{0.5mm} $^b$ Variable identification. 
\hspace*{0.5mm} $^c$ First Fourier amplitude.  
\hspace*{0.5mm} $^d$ Amplitude ratios: $R_{21}$ between second and first 
harmonic, while $R_{31}$ between third and first harmonic.   
\hspace*{0.5mm} $^e$ Phase differences: $\phi_{21}$ between second and 
first harmonic, while $\phi_{31}$ between third and first harmonic.   
\hspace*{0.5mm} $^f$ Reddening evaluations based on the KW relations 
(see text for more details).  
\hspace*{0.5mm} $^g$ Mean (B-V) and (V-I) colors dereddened using the 
KW relations. 
\hspace*{0.5mm} $^h$ Variables that show B and V amplitude modulation. 
} 
\end{deluxetable}

\begin{deluxetable}{lccccc}
\tablecaption{Dereddened colors for the entire sample of RR Lyrae stars} 
\tablehead{
\colhead{ID}&
\colhead{$V_0$\tablenotemark{a}}&
\colhead{$(B-V)_0$\tablenotemark{b}}&
\colhead{$(V-I)_0$\tablenotemark{c}}&
\colhead{$E(B-V)$\tablenotemark{d}} &
\colhead{$E(V-I)$\tablenotemark{d}}
}
\startdata
1         &  13.76   &   0.35   &  0.51   &  0.33    &  0.36\\ 
2$^e$&  13.98   &   0.34   &  0.48  &  0.27   & \ldots \\    
3$^f$   &  13.83   &   0.35   &  0.51  &  0.34    &  0.42\\ 
4         &  13.71   &   0.35   &   0.51  &   0.34   &   0.43\\ 
5         &  13.80   &   0.31   &  0.45  &   0.29    &  0.33\\ 
6$^{e,f}$&  13.74   &   0.31   &   0.45  &  0.30    & \ldots\\ 
7         &  13.82   &   0.38   &   0.54  &  0.27    & \ldots \\ 
8         &  13.71   &   0.38   &  0.54  &  0.32   &   0.29\\ 
9         &  13.81   &   0.33   &  0.48   & 0.31   &   0.30\\ 
11$^g$    &  13.77   &   0.19   &  0.29  & 0.32   &  0.34 \\  
12$^e$ &  13.74   &   0.32   &   0.46  &  0.34   &  \ldots \\ 
13        &  13.75   &   0.35   &   0.50   & 0.35   &   0.32\\ 
17        &  13.90   &   0.35   &   0.50   & 0.27   &   0.34 \\ 
18$^e$ &  13.89   &   0.35   &   0.49   & 0.26   &  \ldots\\ 
19$^f$    &  13.79   &   0.35   &   0.50   & 0.31   &  \ldots \\ 
20$^e$ &  13.78   &   0.33   &   0.48   & 0.30  &   \ldots\\ 
21        &  13.88   &   0.35   &   0.51   & 0.30   &   0.33\\ 
22        &  13.77   &   0.36   &   0.52   & 0.30   &   0.30\\ 
23        &  13.81   &   0.36   &  0.51   &  0.30   &   0.36\\ 
25        &  13.88   &   0.32   &   0.47  &  0.27   &   0.34\\ 
26$^f$    &  13.77   &   0.34   &   0.49  &  0.36   &   0.38\\ 
28        &  13.78   &   0.34   &   0.49  &  0.33   &   0.38\\ 
31$^f$    &  13.99   &   0.34   &   0.50  & 0.30  &    0.41 \\ 
32        &  13.81   &   0.32   &   0.47   & 0.29   &   0.40\\ 
34$^f$    &  13.91   &   0.30   &   0.44  &  0.27   &   0.33\\ 
35        &  13.79   &   0.37   &   0.53  &   0.30   &   0.35\\ 
36        &  14.00   &   0.32   &   0.47   & 0.23   &   0.29\\ 
37$^f$    &  14.01   &   0.39   &   0.56   & 0.24   &   0.26\\ 
38$^{e,f}$& 13.88   &   0.33   &   0.47   & 0.29   &  \ldots\\ 
39        &  13.78   &   0.30   &   0.44   & 0.32   &   0.37\\ 
40        &  13.77   &   0.38   &   0.55   & 0.33   &   0.33\\ 
41        &  13.83   &   0.39   &  0.55   & 0.32   &   0.41\\ 
44        &  13.77   &   0.37   &  0.53   & 0.31   &   0.37\\ 
45        &  13.83   &   0.33   &   0.49   & 0.33  &    0.42\\ 
47$^f$    &  14.04   &   0.37   &   0.52   & 0.20  &    0.19\\                    
48$^{f,g}$  &  13.76   &   0.29   &   \ldots  & 0.30  &  \ldots \\ 
49        &  13.76   &   0.34   &   0.49  & 0.30   &   0.31 \\ 
50$^e$ &  13.86   &   0.36   &   0.47  & 0.24   &  \ldots\\ 
51$^f$    &  13.60   &   0.32   &   0.47  &  0.35   &  \ldots \\ 
56        &  13.81   &   0.36   &   0.51   & 0.35   &   0.44\\ 
57        &  13.92   &   0.36   &   0.52   & 0.30   &  \ldots \\ 
58$^f$    &  13.68   &   0.37   &   0.53   & 0.33   &   0.35\\ 
71        &  13.69   &   0.36   &   0.52   & 0.32   &  \ldots \\ 
73        &  13.69   &   0.31   &   0.46  &  0.34  &    0.36\\ 
76$^e$ &  13.85   &   0.32   &   0.47   & 0.29   & \ldots\\ 
77        &  13.77   &   0.34   &   0.49   & 0.29    &   0.30 \\ 
78        &  13.89   &   0.34   &   0.49   & 0.29    &   0.31\\ 
80        &  13.81   &   0.35   &   0.50  &  0.30    &   0.39\\ 
83        &  13.89   &   0.33   &   0.49  &  0.29    &   0.40\\ 
90        &  13.75   &   0.35   &   0.51   & 0.29    &   0.31\\ 
92$^e$ &  13.77   &   0.33   &   0.48  & 0.30     & \ldots\\  
1405$^g$  &  13.85   &   0.28   &   0.36  &  0.29    &  0.36 \\  
2710      &  13.83   &   0.34   &   0.49  & 0.29     &   0.33\\
\enddata
\tablenotetext{a}{Dereddened mean intensity-weighted V magnitude. The adopted 
reddening correction is the average between the PACZ and the KW relation.
\hspace*{0.5mm} $^b$ Dereddened mean (B-V) color according to PACZ and KW 
relation.
\hspace*{0.5mm} $^c$ Dereddened mean (V-I) color according to PAC and KW 
relation.
rhspace*{0.5mm} $^d$ Reddening correction based on current and KW relations
(weighted mean).
\hspace*{0.5mm} $^e$ Variables whose B and V light curves were fitted 
using the Layden's template (Layden 1998).   
\hspace*{0.5mm} $^f$ Variables that show B and V amplitude modulation.  
\hspace*{0.5mm} $^g$ $RR_c$ pulsator for which the reddening was derived 
using the reddening map of $RR_{ab}$ star.
}
\end{deluxetable}